\documentstyle[amsfonts,amssymb,11pt,epsfig]{article}

\makeatletter
\@addtoreset{equation}{section}

\newcommand{\R}{{\Bbb R}}
\newcommand{\C}{{\Bbb C}}

\def\SL{{\rm SL}}

\def\ul#1{\rlap{\lower1ex\hbox{$-$}}{#1}}

\newcommand{\be}{\begin{eqnarray}}
\newcommand{\ee}{\end{eqnarray}}

\begin{document}

\begin{titlepage}

\thispagestyle{empty}

\title{2D Conformal Field Theories and Holography}

\author{
Laurent {\sc Freidel}\thanks{{\tt
freidel@ens-lyon.fr}}$\hspace{2mm}{}^a$
and
Kirill {\sc Krasnov}\thanks{{\tt
krasnov@cosmic.physics.ucsb.edu}}$\hspace{2mm}{}^b$
\\[10pt]
${}^a${\it Laboratoire de Physique}  \\
{\it Ecole Normale Sup\'erieure de Lyon} \\
{\it 46, all\'ee d'Italie, 69364 Lyon Cedex 07, France}\thanks{{\tt UMR
5672  du CNRS}}\\
{\it and} \\
{\it Perimeter Institute for Theoretical Physics} \\
{\it 35 King Street North} \\
{\it Waterloo N2J 2G9, Ontario, Canada} \\
${}^b${\it Department of Physics, }\\
{\it University of California, Santa Barbara, CA 93106}\\
}

\date{\normalsize May 2002}
\maketitle

\begin{abstract}
\normalsize It is known that the chiral part of any 2d conformal
field theory defines a 3d topological quantum field theory:
quantum states of this TQFT are the CFT conformal blocks. The main aim of
this paper is to show that a similar CFT/TQFT relation exists also for the
full CFT. The 3d topological theory that arises is a certain
``square'' of the chiral TQFT. Such topological theories were
studied by Turaev and Viro; they are related to 3d gravity.
We establish an operator/state correspondence in which operators
in the chiral TQFT correspond to states in the Turaev-Viro theory.
We use this correspondence to interpret CFT correlation functions 
as particular quantum states of the Turaev-Viro theory. 
We compute the components of these states in the basis in the Turaev-Viro 
Hilbert space given by colored 3-valent graphs. The formula we
obtain is a generalization of the Verlinde formula. The later is
obtained from our expression for a zero colored graph. Our results give an 
interesting ``holographic'' perspective on conformal field
theories in 2 dimensions.
\end{abstract}

\end{titlepage}

\section{Introduction}
\label{sec:intr}

To put results of this paper is a somewhat general
context we recall that any conformal field theory (CFT) defines
a topological quantum field theory (TQFT), see \cite{FK} and, e.g., 
\cite{Schweigert} for a review that emphasizes this point. 
The TQFT arises by extracting a
modular tensor category from the CFT chiral
vertex operator algebra. Then, as explained in \cite{Turaev-Book},
any modular category gives rise to a 3d TQFT.
The TQFT can be (partially) described
by saying that its Hilbert space is the the space of (holomorphic)
conformal blocks of the CFT. The canonical example of such
CFT/TQFT correspondence is the well-known relation
between Wess-Zumino-Witten (WZW) and Chern-Simons (CS) theories.
Let us emphasize that this is always a relation between the holomorphic
sector of the CFT (or its chiral part) and a TQFT. As such
it is not an example of a holographic correspondence, in which
correlation functions (comprising both the holomorphic and anti-holomorphic
sectors) of CFT on the boundary would be reproduced by some theory in bulk.

It is then natural to ask whether there is some 3d theory
that corresponds to the {\it full} CFT. A
proposal along these lines was put forward some
time ago by H. Verlinde, see \cite{Verlinde-H}, who argued
that a relation must exist between the quantum Liouville theory
(full, not just the chiral part) and 3d gravity. Recently one
of us presented \cite{Teich} some additional arguments in favor of
this relation, hopefully somewhat clarifying the picture.
The main goal of the present paper is to demonstrate that
such a relation between the full CFT and a certain 3D theory
exists for any CFT. Namely, we show that given a CFT there is a
certain 3d field theory, which is a TQFT, and which is a rather
natural spin-off of the corresponding ``chiral'' TQFT.
The TQFT in question is not new, it is the one defined by Turaev-Viro
\cite{TV}, and described in great detail in \cite{Turaev-Book}.
This paper is thus aimed at a clarification of the relation
between the Turaev-Viro (TV) 3d TQFT's and CFT's in 2 dimensions.

The point that given a CFT there exists a relation between the
full CFT and some 3d TQFT is to some extent contained in recent
works on boundary conformal field theory, see \cite{Schweigert,Gawedzki}
and references therein, and also a more recent paper \cite{Schw-1}. 
As is emphasized, e.g., in \cite{Schweigert}, the full CFT partition 
function on
some Riemann surface $X$ (possibly with a boundary) is equal to
the chiral CFT partition function on the double $\tilde{X}$. There is
then a certain ``connecting'' 3d manifold $\tilde{M}$ whose boundary
$\partial\tilde{M}$ is the double $\tilde{X}$. Using the chiral CFT/TQFT
relation one obtains a 3d TQFT in $\tilde{M}$ that reproduces the
chiral partition function on $\tilde{X}$, and thus the full
partition function on $X$. This formalism turns out to be
very useful for analyzing the case when $X$ has a boundary.

Our analysis was motivated by the above picture, but the logic is
somewhat different. Instead of working with the chiral TQFT in
the connecting 3-manifold $\tilde{M}$ we work directly with
a 3-manifold $M$ whose boundary is $X$, and the Turaev-Viro
TQFT on $M$. The two approaches are equivalent as the TV
theory is a ``square'' of the chiral TQFT. However,
bringing the Turaev-Viro TQFT into the game suggests some
new interpretations and provides new relations.
Thus, most notably, we establish an operator/state correspondence
in which the chiral TQFT operators correspond to states in the 
TV theory, and the trace of an operator product corresponds to the
TV inner product. We use this to interpret the CFT correlators
as quantum states of TV theory. Then, using the fact that
a basis in the Hilbert space of TV theory on $X$ is
given by colored tri-valent graph states,
we will characterize the CFT correlation functions
by finding their components in this basis. Thus, 
the relation that we demonstrate is about a 3d
TQFT on a 3-manifold $M$ and a CFT on the boundary $X$ of $M$.
It is therefore an example of a holographic correspondence, while this
is not obviously so for the correspondence based on a chiral TQFT in
the connecting manifold $\tilde{M}$. 

The holography discussed may be viewed by some as trivial, because the
3-dimensional theory is topological. What makes it interesting
is that it provides a very large
class of examples. Indeed, there is a relation of this type for any CFT.
Importantly, this holography is not limited to any AdS type
background, although a very interesting sub-class of examples
(not considered in this paper, but see \cite{Teich}) is exactly of this type.

As the relation chiral CFT/TQFT is best understood for the case
of a rational CFT, we shall restrict our analysis to this case.
Our constructions can also be expected to generalize to
non-rational and even non-compact CFT's with a continuous
spectrum, but such a generalization is non-trivial, and is
not attempted in this paper. Even with non-compact CFT's excluded,
the class of CFT's that is covered by our considerations, namely,
rational CFT, is still very large. To describe the arising structure 
in its full generality we would need to introduce the apparatus of 
category theory, as it was done, e.g., in \cite{Turaev-Book}. In
order to make the exposition as accessible as possible
we shall not maintain the full generality. We demonstrate the
CFT/TQFT holographic relation using a compact group WZW CFT
(and CS theory as the corresponding chiral TQFT) as an example.

We shall often refer to the TV TQFT as ``gravity''. For
the case of chiral TQFT being the Chern-Simons theory for a group
$G={\rm SU}(2)$ this ``gravity'' theory is just the usual 3d Euclidean
gravity with positive cosmological constant. However, the theory can
be associated to any CFT. The reader should keep in mind
its rather general character.

In order to describe the holographic correspondence in detail we will
need to review (and clarify) the relation
between CS theory and gravity (or between the Reshetikhin-Turaev-Witten
and Turaev-Viro invariants) for a 3-manifold with boundary. We found the
expositions of this relation available in the literature, see
\cite{Turaev-Book,BP}, rather brief and sketchy. This paper provides
a more detailed account and obtains new results. In particular, 
the operator/state correspondence established in this paper is new.

The paper is organized as follows.
In section \ref{sec:CS} we review the quantization of Chern-Simons theory.
Section \ref{sec:TV} is devoted to the Turaev-Viro theory.
We then review the definition of 3-manifold invariants in
section \ref{sec:inv}, and some facts on the Verlinde formula in
section \ref{sec:ver}. The new material starts in section
\ref{sec:rel}, where we discuss the CS/TV operator/state
correspondence and the arising relation between the CS and TV Hilbert spaces. 
In section \ref{sec:part} we  interpret
the CFT partition function as a TV quantum state, and
compute components of this state in a natural basis in the
TV Hilbert space given by graphs. We conclude with a discussion.

\section{Chern-Simons theory}
\label{sec:CS}

This section is a rather standard review of CS theory. We discuss
the CS phase space, the Hilbert space that arises as its quantization,
review the Verlinde formula, and a particular basis in the
CS Hilbert space that arises from a pant decomposition.
The reader may consult, e.g., \cite{Witten-Jones,Turaev-Book} for
more details.

\bigskip
\noindent{\bf Action.} The Chern-Simons (CS) theory is
a 3-dimensional TQFT of Witten type. The CS theory for a
group $G$ is defined by the following action functional:
\be\label{cs-action}
S_{\rm CS}^-[A] = {k\over 4\pi} \int_M {\rm Tr} \left( A\wedge dA +
{2\over 3} A\wedge A\wedge A\right) -{k\over 4\pi} \int_{\partial M}
dz\wedge d\bar{z}\,\,{\rm Tr}(A_z A_{\bar{z}}).
\ee
Here $M$ is a 3-dimensional manifold and $A$ is a connection on
the principal $G$-bundle over $M$.
For the case of a compact $G$ that we consider in this paper the action
is gauge invariant (modulo $2\pi$) when $k$ is an integer. The second term
in (\ref{cs-action}) is necessary to make the action principle
well-defined on a manifold with boundary. To write it one
needs to choose a complex structure on $\partial M$. The term in
(\ref{cs-action}) is the one relevant for
fixing $A_{\bar{z}}$ on the boundary. Another possible choice of boundary
condition is to fix $A_z$. The corresponding action is:
\be\label{cs-action'}
S_{\rm CS}^+[A] = {k\over 4\pi} \int_M {\rm Tr} \left( A\wedge dA +
{2\over 3} A\wedge A\wedge A\right) +{k\over 4\pi} \int_{\partial M}
dz\wedge d\bar{z}\,\,{\rm Tr}(A_z A_{\bar{z}}).
\ee

\bigskip
\noindent{\bf Partition function.} The partition function arises
(formally) by considering the path integral for (\ref{cs-action}).
For a closed $M$ it can be given a precise meaning through the
surgery representation of $M$ and the Reshetikhin-Turaev-Witten
(RTW) invariant of links. Before we review this
construction, let us discuss the formal path integral for the
case when $M$ has a boundary. For example, 
let the manifold $M$ be a handlebody $H$.
Its boundary $X=\partial H$ is a (connected) Riemann surface.
Recall that TQFT assigns a Hilbert space to each connected component of
$\partial M$, and a map between these Hilbert spaces (functor) to
$M$. The map can be heuristically thought of as given by the
path integral. For a manifold with a single boundary component,
which is the case for a handlebody $H$, TQFT on $H$ gives a functor
${\cal F}: {\cal H}^{\rm CS}_X \to \C$ mapping the CS Hilbert space
of $X$ into $\C$. This functor can be obtained from the following
Hartle-Hawking (HH) type state:
\be\label{cs-path-int}
{\cal F}(\ul{A})=\int_{A_{\bar{z}}=\ul{A}} {\cal D}A\,\, e^{i S_{\rm CS}^-[A]}.
\ee
The path integral is taken over connections in $H$ with
the restriction of $A$ on $X$ fixed. More precisely, with
the choice of boundary term in the action as in (\ref{cs-action}),
one fixes only the anti-holomorphic part $\ul{A}=A_{\bar{z}}$ of the
connection on $X$, as defined by an auxiliary
complex structure. The result of the path integral
(\ref{cs-path-int}) is the partition function of CS theory on $H$. It can
be thought of as a particular quantum state ${\cal F}(\ul{A})$ in the
CS Hilbert space ${\cal H}^{\rm CS}_X$. The inner product in
${\cal H}^{\rm CS}_X$ is (formally) defined as:
\be\label{cs-inner-prod}
\langle \Psi_1 \mid \Psi_2 \rangle = {1\over {\rm Vol}\,{\cal G}}
\int_{\cal A} {\cal D}\ul{A}\overline{{\cal D}\ul{A}}\,\, 
\overline{\Psi_1(\ul{A})} \Psi_2(\ul{A}).
\ee
Since the integrand is gauge invariant it is natural to divide
by the volume ${\rm Vol}\,{\cal G}$ of the group of gauge
transformations. The above mentioned functor
${\cal F}: {\cal H}^{\rm CS}_X \to \C$ is given by:
\be
{\cal F}(\Psi)=\langle {\cal F} \mid \Psi \rangle ={1\over {\rm Vol}\,{\cal G}}
\int_{\cal A} {\cal D}\ul{A}\overline{{\cal D}\ul{A}}\,\, 
\overline{{\cal F}(\ul{A})} \Psi(\ul{A}).
\ee
The state ${\cal F}(\ul{A})\in{\cal H}^{\rm CS}_X$ depends only
on the topological nature of the 3-manifold and a framing of $M$.

\bigskip
\noindent{\bf Phase space.} To understand the
structure of the CS Hilbert space ${\cal H}^{\rm CS}$ it is
natural to use the Hamiltonian description. Namely,
near the boundary the manifold has the topology $X\times\R$.
Then the phase space ${\cal P}^{\rm CS}$ of CS theory based on a
group $G$ is the moduli space of flat $G$-connections on $X$
modulo gauge transformations:
\be
{\cal P}^{\rm CS}_X\sim {\cal A}/{\cal G}.
\ee
It is finite dimensional.

Let $X$ be a (connected) Riemann surface of type $(g,n)$ with
$g\geq 0, n>0, 2g+n-2>0$. Denote the fundamental group of $X$
by $\pi(X)$. The moduli space ${\cal A}$ can then
be parametrized by homomorphisms $\phi: \pi(X)\to G$. The phase
space is, therefore, isomorphic to
\be\label{cs-phase-space}
{\cal P}^{\rm CS}_X\sim {\rm Hom}(\pi(X),G)/G,
\ee
where one mods out by the action of the group at the base point.
The fundamental group is generated by $m_i, i=1,\ldots,n$ and
$a_i, b_i, i=1,\ldots,g$ satisfying the following relation:
\be
m_1\ldots m_n [a_1,b_1]\ldots [a_g,b_g]=1.
\ee
Here $[a,b]=a b a^{-1} b^{-1}$. The dimension of the phase space
can now be seen to be:
\be\label{dim-cs}
{\rm dim}\,{\cal P}^{\rm CS}_X=(2g+n-2){\rm dim}\,G.
\ee
The fact that (\ref{cs-phase-space}) is naturally a Poisson
manifold was emphasized in \cite{Goldman}. The Poisson
structure described in \cite{Goldman} is the same as the one
that comes from CS theory. For the case of a compact $X$
the space (\ref{cs-phase-space}) is actually a symplectic manifold.
For the case when punctures are present the symplectic leaves
are obtained by restricting the holonomy of $\ul{A}$ around
punctures to lie in some conjugacy classes in the group.
An appropriate power of the
symplectic structure can be used as a volume form on
the symplectic leaves. Their volume turns out to be
finite. One thus expects to get finite dimensional Hilbert spaces upon
quantization.

\bigskip
\noindent{\bf Hilbert space.} The Hilbert space ${\cal H}^{\rm CS}_X$
was understood \cite{Witten-Jones,Witten-Holomorphic} to be the
same as the space of conformal blocks of the chiral Wess-Zumino-Witten (WZW)
theory on a genus $g$-surface with $n$ vertex operators inserted.
Let us fix conformal dimensions of the operators inserted, that is,
fix irreducible representations ${\bf R}=\{\rho_1,\ldots,\rho_n\}$
of $G$ labelling the punctures. The dimension of each of ${\cal H}^{\rm CS}_X$
can be computed using the Verlinde formula \cite{Verlinde-E,MS}:
\be\label{ver-formula}
{\rm dim}\,{\cal H}^{\rm CS}_X = \sum_\rho {S_{\rho_1 \rho}\ldots
S_{\rho_n \rho}\over S_{0 \rho}\ldots S_{0 \rho}} (S_{0 \rho})^{2-2g}.
\ee
The sum is taken over irreducible representations $\rho$, 
$S_{\rho \rho'}$ is the modular S-matrix, see (\ref{s-matrix})
below for the case of ${\rm SU}(2)$, and $S_{0\rho}=\eta\, {\rm dim}_\rho$,
where $\eta$ is given by (\ref{eta}).

\bigskip
\noindent{\bf Pant decomposition.} The states
from ${\cal H}^{\rm CS}_X$ can be understood as the
HH type states given by the path integral over a handlebody $H$
with Wilson lines in representations ${\bf R}$
intersecting the boundary $X$ transversally at $n$ points. A
convenient basis in ${\cal H}^{\rm CS}_X$ can be obtained
by choosing a pant decomposition of $X$. A pair of pants
is a sphere with 3 holes (some of them can be punctures).
A Riemann surface $X$ of type $(g,n)$ can be represented by
$2g+n-2$ pants glued together. For example, the surface of
type $(0,4)$ with 4 punctures can be obtained by gluing
together 2 spheres each with 2 punctures and one hole.
Note that a pant decomposition is not unique. Different
pant decompositions are related by simple ``moves''.
A pant decomposition can be conveniently encoded
in a tri-valent graph $\Delta$ with $2g+n-2$ vertices and
$3g+2n-3$ edges. Each vertex of $\Delta$ corresponds
to a pair of pants, and each internal edge corresponds to two holes
glued together. Open-ended edges of $\Delta$ end at punctures.
We shall call such edges ``loose''. There are exactly
$n$ of them. The graph $\Delta$ can be thought
of as a 1-skeleton of the Riemann surface $X$, or as a Feynman
diagram that corresponds to the string world-sheet $X$.
The handlebody $H$ can be obtained from
$\Delta$ as its regular neighborhood $U(\Delta)$, so that $\Delta$ is
inside $H$ and the loose edges of $\Delta$ end at the punctures.
Let us label the loose edges by representations $\bf R$ and internal
edges by some other irreducible representations. It is
convenient to formalize this labelling in a notion of
{\it coloring} $\phi$. A coloring $\phi$ is the map
\be
\phi: E_\Delta\to {\cal I}, \qquad \phi(e)=\rho_e \in {\cal I}
\ee
from the set $E_\Delta$ of edges of $\Delta$ to the set
${\cal I}$ of irreducible representations of the quantum group $G$. The
loose edges are colored by representations from $\bf R$.
The CS path integral on $H$ with the spin network $\Delta^\phi$
inserted is a state in ${\cal H}^{\rm CS}_X$. See below for a definition
of spin networks. Changing the labels on the internal edges one
gets states that span the whole ${\cal H}^{\rm CS}_X$. Different
choices of pant decomposition of $H$ (and thus of $\Delta$) lead
to different bases in ${\cal H}^{\rm CS}_X$.

\bigskip
\noindent{\bf Inner product.} The inner product
(\ref{cs-inner-prod}) of two states of the type described can be obtained
by the following operation. Let one state be given by the path integral
over $H$ with $\Delta^\phi$ inserted and the other by $H$ with
${\Delta^\phi}'$ inserted, where
both the graph and/or the coloring may be different in the two states.
Let us invert orientation of the first copy of $H$ and glue $-H$
to $H$ across the boundary (using the identity homomorphism)
to obtain some 3d space $\tilde{H}$ without boundary. We will refer to
$\tilde{H}$ as the {\it double} of $H$. For $H$ being
a handlebody with $g$ handles the double $\tilde{H}$ has the topology
of a connected sum:
\be
\tilde{H}\sim \#_{g-1} S^2\times S^1.
\ee
The loose ends of $\Delta$ are connected at the punctures to the
loose ends of $\Delta'$ to obtain a colored closed graph
$\Delta^\phi\cup{\Delta^\phi}'$ inside $\tilde{H}$. The inner product
(\ref{cs-inner-prod}) is given by the CS path integral over $\tilde{H}$
with the spin network $\Delta^\phi\cup{\Delta^\phi}'$ inserted.
This path integral is given by the RTW evaluation
of $\Delta^\phi\cup{\Delta^\phi}'$ in $\tilde{H}$, see below for
a definition of the RTW evaluation.

\section{Gravity}
\label{sec:TV}

The material reviewed in this section is less familiar, although is
contained in the literature. We give the
action for Turaev-Viro theory, discuss the phase space, then
introduce certain important graph coordinatization of it,
define spin networks, and describe the TV Hilbert space. A
useful reference for this section is the book of Turaev \cite{Turaev-Book}
and the paper \cite{Baez}.

\bigskip
\noindent{\bf Action.} What we refer to as ``gravity'' arises as a
certain ``square'' of CS theory. We will also refer to this
gravity theory as Turaev-Viro (TV) theory, to have uniform
notations (CS-TV).

To see how the TV theory (gravity) arises from CS theory,
let us introduce two connection fields $A$ and $B$.
Consider the corresponding CS actions $S_{\rm CS}[A], S_{\rm CS}[B]$.
Introduce the following parameterization of the fields:
\be\label{A-B}
A={\bf w}+\left({\pi\over k}\right) {\bf e}, 
\qquad B={\bf w}- \left({\pi\over k}\right) {\bf e}.
\ee
Here $\bf w$ is a $G$-connection, and ${\bf e}$ is a one-form
valued in the Lie algebra of $G$. The TV theory action is 
essentially given by the difference $S_{\rm CS}^-[A]-S_{\rm CS}^+[B]$,
plus a boundary term such that the full action is:
\be\label{tv-action}
S_{\rm TV}[{\bf w},{\bf e}]= \int_M {\rm Tr}\left(
{\bf e}\wedge {\bf f}({\bf w})+ {\Lambda \over 12}
{\bf e}\wedge{\bf e}\wedge{\bf e} \right).
\ee
The boundary condition for this action is that the
restriction $\ul{\bf w}$ of $\bf w$ on $X=\partial M$ is
kept fixed. Here $\Lambda$ is the ``cosmological constant'' related to $k$ as:
$k=2\pi/\sqrt{\Lambda}$. For $G={\rm SU}(2)$ the TV theory is nothing 
else but the Euclidean gravity with
positive cosmological constant $\Lambda$. We emphasize, however, that 
the theory is defined for
other groups as well. Moreover, it also exists
as a square of a chiral TQFT for any TQFT, that is
even in cases when the chiral TQFT is not a CS theory.

\bigskip
\noindent{\bf Path integral.} Similarly to CS theory, one can
consider HH type states given by the path integral on a manifold with
a single boundary component. Thus, for a manifold being a
handlebody $H$ we get the TV partition function:
\be\label{tv-path-int}
{\cal T}(\ul{\bf w})=\int_{{\bf w}\mid_X=\ul{{\bf w}}}
{\cal D}{\bf w}{\cal D}{\bf e}\,\,
e^{i S_{\rm TV}[{\bf w},{\bf e}]}.
\ee
The integral is taken over both $\bf w, e$ fields in the bulk with
the restriction $\ul{\bf w}$ of the connection fixed on the
boundary. The TV partition function ${\cal T}(\ul{\bf w})$ is
thus a functional of the boundary connection. It can
also be interpreted as a particular state in the TV Hilbert space
${\cal H}^{\rm TV}_X$.

States from ${\cal H}^{\rm TV}_X$ are functionals
of the boundary connection. The inner product on this space can
be formally defined by the formula
\be\label{tv-inner-prod}
\langle \Psi_1 \mid \Psi_2 \rangle = {1\over {\rm Vol}\,{\cal G}}
\int_{\cal A} {\cal D}\ul{\bf w}\,\, \overline{\Psi_1(\ul{\bf w})}
\Psi_2(\ul{\bf w})
\ee
similar to (\ref{cs-inner-prod}). Note, however, that the measure in
(\ref{tv-inner-prod}) is different from that in (\ref{cs-inner-prod}).
We shall see this below when we describe how
to compute TV inner products in practice.

\bigskip
\noindent{\bf Phase space.} The TV phase space is
basically two copies of ${\cal P}^{\rm CS}$, but with an
unusual polarization. The polarization on ${\cal P}^{\rm TV}$ 
is given by $\bf e, w$, which are canonically conjugate variables. 
Note that there is no need to choose a complex structure in order 
to define this polarization.

It turns out to be very convenient to think of ${\cal P}^{\rm TV}$ as
some deformation of the cotangent bundle $T^*({\cal A}/{\cal G})$
over the moduli space ${\cal A}/{\cal G}$ of flat connections on $X$.
Note, however, that the TV connection $\ul{\bf w}$ on the boundary
is not flat, so the configuration space for TV theory is not really
the moduli space of flat connections. One does get ${\cal A}/{\cal G}$
as the configurational space in an important limit $k\to\infty$, in
which the ${\bf e}^{\wedge 3}$ term drops from the action (\ref{tv-action}).
Thus, it is only in this limit that the TV phase space is the
cotangent bundle $T^*({\cal A}/{\cal G})$. For a finite $k$
the TV phase space is compact (as consisting of two copies of
${\cal P}^{\rm CS}$), while $T^*({\cal A}/{\cal G})$ is not.
We will see, however, that it is essentially correct to think
of ${\cal P}^{\rm TV}$ as a deformation of $T^*({\cal A}/{\cal G})$ even in
the finite $k$ case. The compactness of ${\cal P}^{\rm TV}$
will manifest itself in the fact that after the quantization 
the range of eigenvalues of $\bf e$ is bounded from above.\footnote{%
An interesting analogy was suggested to one of us by R.\ Roiban. The
analogy is with Calabi-Yau manifolds that are compact, but whose
topology near a special Lagrangian submanifold is locally that
of a cotangent bundle. We don't know whether this is
just an analogy or ${\cal P}^{\rm TV}$ is indeed a Calabi-Yau
space.}

These remarks being made we write:
\be
{\cal P}^{\rm TV}\sim T^*_k \left({\cal A}/{\cal G}\right),
\ee
where $T^*_k$ is certain compact version of
the cotangent bundle. The phase space becomes the usual cotangent
bundle in the $k\to\infty$ limit. We will not need any further
details on spaces $T^*_k$. As we shall see the quantization
of ${\cal P}^{\rm TV}$ is rather straightforward once the
quantization of the cotangent bundle is understood.

We note that the dimension
\be
{\rm dim}\,{\cal P}^{\rm TV}=2(2g+n-2){\rm dim}\, G
\ee
is twice the dimension of the phase space of the corresponding
CS theory, as required. A convenient parameterization of the
cotangent bundle phase space can be obtained by using graphs.

\begin{figure}
\centerline{\hbox{\epsfig{figure=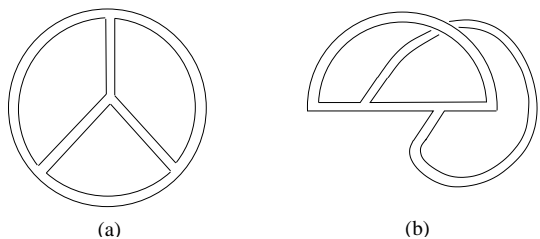,height=1.2in}}}
\caption{A fat graph $\Gamma$ for the: (a) sphere with
4 punctures; (b) torus with one puncture.}
\label{fig:fat}
\end{figure}

\bigskip
\noindent{\bf Graphs.}
The graphs one considers are similar to those that arise in the
Penner coordinatization \cite{Penner} of the moduli space of
punctured Riemann surfaces. Namely, given $X$, introduce a tri-valent
closed fat graph $\Gamma$ with the number $F$ of faces equal to
the number $n$ of punctures. Such a graph can be obtained by
triangulating the surface $X$ using punctures as vertices, and
then constructing a dual graph. What arises is exactly a graph $\Gamma$.
See Fig.~\ref{fig:fat} for examples of $\Gamma$.
Note that different triangulations lead to different graphs, so
$\Gamma$ is by no means unique.

Because the graph is tri-valent $3V=2E$, where
$V$ is the number of vertices and $E$ is the number of edges.
We also have the Euler characteristics relation:
\be
F-E+V=2-2g.
\ee
We thus get that the number $E$ of edges of $\Gamma$ is
$E=3(2g+n-2)$.

Note that the graph $\Gamma$
does not coincide with the graph $\Delta$ introduced in the
previous section. There is, however, a simple relation between them
that is worth noting. Let us, as in the previous section, form
the double $\tilde{H}=H\cup -H$. It is a closed 3-manifold obtained
by gluing two copies of the handlebody $H$ across the boundary $X$.
Let us take a graph $\Delta$ in $H$, and another copy of
$\Delta$ in $-H$. These graphs touch the boundary $\partial H=X$
at the punctures. Gluing these two copies of $\Delta$ at the
punctures one obtains a closed graph $\Delta\cup\Delta$ in
$\tilde{H}$. It is a tri-valent graph with $2(2g+n-2)$ vertices
and $3(2g+n-2)$ edges. Now consider the regular
neighborhood $U(\Delta\cup\Delta)$ of $\Delta\cup\Delta$ in
$H\cup -H$. This is a handlebody, whose boundary is of
genus
\be\label{G}
G=2g+n-1.
\ee
The surface $\partial U(\Delta\cup\Delta)$ can be obtained
by taking two copies of $X$, removing some small disks
around the punctures, and identifying the resulting circular
boundaries to get a closed surface without punctures. We have the following

\bigskip
\noindent{\bf Lemma.} {\it The surface $\partial U(\Delta\cup\Delta)$
is a Heegard surface for $H\cup -H$. The complement of
$U(\Delta\cup\Delta)$ in $H\cup -H$ is a handlebody that
is the regular neighborhood $U(\Gamma)$ of
the graph $\Gamma$ on $X$.}

\bigskip
\noindent{\it Proof.} The complement of
$U(\Delta\cup\Delta)$ in $H\cup -H$ can be seen to be
the cylinder $X\times [0,1]$ with $n$ holes cut in it. So,
it is indeed a handlebody of genus (\ref{G}). Its 1-skeleton
that can be obtained by choosing a pant decomposition is
the tri-valent graph $\Gamma$.

\bigskip
\noindent{\bf Graph connections.}
Given $\Gamma$ equipped with an arbitrary orientation of all the edges,
one can introduce what can be called graph
connections. Denote the set of edges $e$ of
$\Gamma$ by $E$. We use the same letter both for the set $E$ of edges
and for its dimension.
A graph connection $\bf A$ is an assignment of a group
element to every edge of the graph:
\be
{\bf A}: E\to G, \qquad {\bf A}(e)={\bf g}_e \in G.
\ee
One can also introduce a notion of graph gauge transformations. These
act at vertices of $\Gamma$. A gauge transformation is parameterized
by $V$ group elements. Let us introduce:
\be\label{graph-gauge-tr}
{\bf H}: V\to G, \qquad {\bf H}(v)={\bf h}_v \in G.
\ee
Here $V$ is the set of vertices of $\Gamma$. For an edge $e\in E$ denote by
$s(e)$ (source) the vertex from which $e$ originate, and
by $t(e)$ (target) the vertex where $e$ ends. The action of a gauge
transformation $\bf H$ on a graph connection $\bf A$ is now
as follows:
\be
{\bf A}^{\bf H}(e) = {\bf h}^{-1}_{s(t)}\, {\bf g}_e \,{\bf h}_{s(e)}.
\ee
The space of graph connections modulo graph gauge transformations
can now be seen to be isomorphic to $G^{\otimes E}/G^{\otimes V}$. Its
dimension is given by (\ref{dim-cs}). We thus get
a parameterization of the CS phase space ${\cal P}^{\rm CS}$ based on
a graph $\Gamma$:
\be\label{cs-phase-space-graph}
{\cal P}^{\rm CS}\sim G^{\otimes E}/G^{\otimes V}.
\ee
The TV phase space is the cotangent bundle
\be\label{tv-phase-space-graph}
{\cal P}^{\rm TV} \sim T^*_k \left(G^{\otimes E}/G^{\otimes V}\right).
\ee
As we shall see, it is rather straightforward to quantize the
non-compact, $k\to\infty$ version of ${\cal P}^{\rm TV}$,
that is the cotangent bundle. The quantum states are given by
spin networks.

\bigskip
\noindent{\bf Spin networks.} To quantize the cotangent
bundle $T^*({\cal A}/{\cal G})$ one introduces a Hilbert space
of functionals on the moduli space of flat connections.
A complete set of such functionals is given by spin networks.
These functions will thus form (an over-complete) basis in 
the Hilbert space of TV theory. They also serve as 
observables for CS quantum theory, see below.

Before we define these objects, let us introduce some convenient notations.
Denote the set of irreducible representations $\rho$ of the
quantum group $G$ by $\cal I$.
Introduce a coloring $\psi:E\to {\cal I}, \psi(e)=\rho_e$ of
the edges of $\Gamma$ with irreducible representations of $G$. A
spin network $\Gamma^\psi$ is a functional on the space of graph
connections:
\be\label{spin-net}
\Gamma^\psi: G^{\otimes E}\to \C.
\ee
Given a connection $\bf A$ the value of $\Gamma^\psi({\bf A})$
is computed as follows. For every edge $e$ take the group
element ${\bf g}_e$ given by the graph connection in the
irreducible representation $\rho_e$. One can think of this as a
matrix with two indices: one for the source $s(e)$ and the other
for the target $t(e)$. Multiply the matrices for all the edges of
$\Gamma$. Then contract the indices at every tri-valent
vertex using an intertwining operator. The normalization of intertwiners 
that we use is specified in the Appendix. We assume,
for simplicity, that the group $G$ is such that the tri-valent
intertwiner is unique. An example is given by ${\rm SU}(2)$.
If this is not so, one should in addition
label the vertices of $\Gamma$ with intertwiners, so that a
spin network explicitly depends on this labelling. The
functional (\ref{spin-net}) so constructed is invariant
under the graph gauge transformations (\ref{graph-gauge-tr})
and is thus a functional on the moduli space of flat connections
modulo gauge transformations. As such it is an element of the
Hilbert space of TV theory. It is also an observable on the CS phase space
(\ref{cs-phase-space-graph}).

\bigskip
\noindent{\bf Quantization.} We can define
the Hilbert space ${\cal H}^{\rm TV}$ of Turaev-Viro theory
to be the space of gauge-invariant functionals
$\Psi(\ul{\bf w})$ on the configurational
space $G^{\otimes E}/G^{\otimes V}$. This gives a quantization
of the $k\to\infty$ limit, but a modification for the case of
finite $k$ is straightforward. As we discussed above,
a complete set of functionals on
$G^{\otimes E}/G^{\otimes V}$ is given by spin networks. We denote the state
corresponding to a spin network $\Gamma^\psi$ by $|\Gamma^\psi\rangle$.
They form a basis of states in ${\cal H}^{\rm TV}$:
\be
{\cal H}^{\rm TV}={\rm Span}\{ \mid \Gamma^\psi\rangle \}.
\ee
One can construct certain momenta operators, analogs of
${\bf e}\sim \partial/\partial \ul{\bf w}$
in the continuum theory. Spin networks are eigenfunctions of
these momenta operators. To specialize to the case of finite
$k$ one has to replace all spin networks by quantum ones.
That is, the coloring of edges of $\Gamma$ must use
irreducible representations of the quantum group,
which there is only a finite set.

The spin network states $\mid \Gamma^\psi\rangle$ form an
over-complete basis in ${\cal H}^{\rm TV}$, in that the
TV inner product between differently colored states 
is non-zero. However, these states do give a partition of unity
in that
\be\label{identity-tv}
\sum_\psi \left( \prod_{e\in E_\Gamma} {\rm dim}_{\rho_e} \right)
\mid \Gamma^\psi\rangle \langle \Gamma^\psi \mid
\ee
is the identity operator in ${\cal H}^{\rm TV}$. This will
become clear from our definition of the TV inner product, and
the definition of the TV invariant in the next section.

It seems from the way we have constructed the Hilbert space
${\cal H}^{\rm TV}$ that it depends on the graph $\Gamma$. This
is not so. Choosing $\Gamma$ differently one gets a different
basis in the same Hilbert space. To describe an effect of
a change of $\Gamma$ it is enough to give a rule for
determining the inner products between states from two
different bases.

\bigskip
\noindent{\bf Inner product.} The inner product on ${\cal H}^{\rm TV}$
is given (formally) by the integral (\ref{tv-inner-prod})
over boundary connections. To specify the measure in this
integral, one has to consider the path integral for the theory.
Namely, consider a 3-manifold $X\times [-1,1]$ over $X$,
which is a 3-manifold with two boundary components, each of which
is a copy of $X$. The TV path integral over $X\times [-1,1]$
gives a kernel that should be sandwiched between the two
states whose inner product is to be computed. Thus, the
measure in (\ref{tv-inner-prod}) is defined by the TV path integral.
The measure, in particular, depends on the level $k$.

In practice the inner product of two states $\Gamma^\psi$
${\Gamma^\psi}'$, where both the graphs and the coloring may
be different, is computed as the TV invariant, see below,
for the manifold $X\times [-1,1]$ with $\Gamma^\psi$ on
$X\times \{-1\}$ and ${\Gamma^\psi}'$ on $X\times \{1\}$.

\section{3-Manifold invariants}
\label{sec:inv}

In this section we review the definition of RTW and
TV invariants. The main references for this section are 
\cite{Roberts} and \cite{Turaev}. 

\bigskip
\noindent{\bf Reshetikhin-Turaev-Witten invariant.} The RTW invariant
of a closed 3-manifold (with, possibly, Wilson loops or spin networks inserted)
gives a precise meaning to the CS path integral for this manifold.
The definition we give is for $M$ without insertions, and is different 
from, but equivalent to the original definition in \cite{RT}.
We follow Roberts \cite{Roberts}.

Any closed oriented 3-manifold $M$ can be obtained from $S^3$ by
a surgery on a link in $S^3$. Two framed links represent the
same manifold $M$ if and only if they are related by isotopy or a sequence
of Kirby moves, that is either handle-slides or blow-ups, see
\cite{Roberts} or \cite{KL} for more detail. Let $L$ be a link giving
the surgery representation
of $M$. Define $\Omega L\in\C$ to be the evaluation of $L$ in
$S^3$ with a certain element $\Omega$ inserted along all
the components of $L$, paying attention to the framing.
The element $\Omega$ is defined as follows, see \cite{Roberts}. It is
an element of ${\cal H}^{\rm CS}_{T}$, where $T$ is the torus, and
is given by:
\be\label{omega}
\Omega=\eta \sum_\rho {\rm dim}_\rho R_\rho.
\ee
The sum is taken over all irreducible representations $\rho\in{\cal I}$,
the quantity ${\rm dim}_\rho$ is the quantum dimension, and $R_\rho$ is
the state in ${\cal H}^{\rm CS}_T$ obtained by inserting the 0-framed
unknot in the $\rho$'s representation along the cycle that
is non-contractible inside the solid torus having $T$ as its boundary.
The quantity $\eta$ is given by:
\be\label{eta}
\eta^{-2}=\sum_\rho {\rm dim}_\rho^2.
\ee
For example, for $G={\rm SU}(2)$ $\eta=\sqrt{2/k}\,\sin(\pi/k)=S_{00}$,
where
\be\label{s-matrix}
S_{ij}=\sqrt{2\over k} \sin\left( {(i+1)(j+1)\pi\over k}\right),
\qquad k\geq 3.
\ee
With the normalization chosen, the $S^3$ value of a 0-framed unknot with
$\Omega$ attached is $\eta^{-1}$, while $\pm 1$ framed unknots with
$\Omega$ attached give certain unit modulus complex numbers $r^{\pm 1}$.
For $G={\rm SU}(2)$ $r=\exp(-i\pi/4-2\pi i(3+k^2)/4k)$.

Let us now continue with the definition of the RTW invariant. Define
by $\sigma(L)$ the signature of the 4-manifold obtained by attaching
2-handles to the 4-ball $B^4$ along $L\subset S^3=\partial B^4$. Define
\be
I(M)=\eta r^{-\sigma(L)}\, \Omega L.
\ee
This is the RTW invariant of the manifold $M$ presented by $L$. We use
the normalization of Roberts \cite{Roberts}, in which the RTW invariant
satisfies $I(S^3)=\eta, I(S^2\times S^1)=1$, as well as the connected
sum rule $I(M_1 \# M_2)=\eta^{-1} I(M_1) I(M_2)$.

\bigskip
\noindent{\bf Turaev-Viro invariant.} The original Turaev-Viro
invariant is defined \cite{TV} for triangulated manifolds. A
more convenient presentation \cite{Turaev} uses standard 2-polyhedra.
Another definition is that of Roberts \cite{Roberts}. It uses a
handle decomposition of $M$. We first give the original definition of 
Turaev and Viro.

Let $T$ be a triangulation of 3d manifold $M$. We are mostly interested
in case that $M$ has a boundary. Denote by $V_T$ the number of vertices
of $T$, and by $\{ e\}, \{ f\}, \{t\}$ collections of edges, faces
and tetrahedra of $T$. Choose a coloring $\mu$ of all the edges, so
that $\mu(e)=\rho_e$ is
the color assigned to an edge $e$. The Turaev-Viro invariant
is defined as:
\be\label{tv-triang}
{\rm TV}(M,T|_{\partial M},\mu|_{\partial M})=
\eta^{2V_T} \sum_\mu \prod_{e\notin\partial M}
{\rm dim}_{\rho_e} \prod_t
(6j)_t.
\ee
Here $(6j)_t$ is the 6j-symbol constructed out of 6 colors
labelling the edges of a tetrahedron $t$, and the product is taken
over all tetrahedra $t$ of $T$. The product of dimensions of
representations labelling the edges is taken over all edges that
do not lie on the boundary. The sum is taken over all colorings $\mu$ keeping
the coloring on the boundary fixed. The invariant depends on the
restriction $T|_{\partial M}$ of the triangulation to the
boundary $\partial M$, and on the coloring $\mu|_{\partial M}$ of
this restriction. The invariant is independent of
an extension of $T|_{\partial M}$ inside $M$.

Note that the TV invariant is constructed in such a way that
for a closed 3-manifold $M=M_1\cup M_2$ obtained by gluing
two manifolds $M_1, M_2$ with a boundary across the boundary
the invariant $TV(M)$ is easily obtained once 
$TV(M_{1,2}, T_{1,2}|_{\partial M_{1,2}},\mu_{1,2}|_{\partial M_{1,2}})$
are known. One has to triangulate the boundary of $M_{1,2}$ in the
same way $T_1|_{\partial M_1}=T_2|_{\partial M_2}$, multiply the 
invariants for $M_{1,2}$, 
multiply the result by the dimensions of the representations 
labelling the edges of $T_{1,2}|_{\partial M_{1,2}}=T|_{\partial M}$, 
and sum over these representations. The result is $TV(M)$:
\be
TV(M)=\sum_{\mu|_{\partial M}} \left( \prod_{e\in\partial M} 
{\rm dim}_{\rho_e}\right)
TV(M_1, T|_{\partial M},\mu|_{\partial M})
TV(M_2, T|_{\partial M},\mu|_{\partial M}).
\ee
This, together with the definition of the TV inner product
as the TV invariant for $X\times I$ establishes that 
(\ref{identity-tv}) is indeed the identity operator in
${\cal H}^{\rm TV}$.

\bigskip
\noindent{\bf Roberts invariant.} We shall now introduce the more
general invariant of Roberts. We consider the case without boundary.

Consider a handle decomposition $D$ of $M$. The canonical example to
have in mind is the handle decomposition coming from a triangulation
$T$ of $M$. A thickening of the corresponding dual complex
$T^*$ then gives a handle decomposition.
The vertices of the dual complex (baricenters of tetrahedra of the
triangulation) correspond to 0-handles, edges of $T^*$
(faces of $T$) correspond to 1-handles, faces of
$T^*$ (edges of $T$) give 2-handles,
and 3-cells of $T^*$ (vertices of $T$)
give 3-handles. The union of 0- and 1-handles is a handlebody.
Choose a system of meridian discs for it, one
meridian discs for every 1-handle. Now specify the system of attaching curves
for 2-handles. If the handle decomposition came from a
triangulation there are exactly 3 attaching curves along
each 1-handle. Frame all meridian and attaching curves using
the orientation of the boundary of the handlebody.
Denote the corresponding link by $C(M,D)$.
Insert the element $\Omega$ on all the components of $C(M,D)$,
paying attention to the framing, and evaluate $C(M,D)$
in $S^3$. This gives the Roberts invariant for $M$:
\be\label{r-inv}
{\rm R}\,(M)= \eta^{d_3+d_0} \Omega C(M,D).
\ee
Here $d_3, d_0$ are the numbers of 3- and 0-handles correspondingly.
Note that to evaluate $\Omega C(M,D)$ in $S^3$ one needs to
first specify an embedding. The result of the evaluation does not
depend on the embedding, see \cite{Roberts}. Moreover, the
invariant does not depends on a handle decomposition $D$ and
is thus a true invariant of $M$.

When the handle
decomposition $D$ comes from a triangulation $T$ the Roberts
invariant (\ref{r-inv}) coincides with the Turaev-Viro invariant
(\ref{tv-triang}). An illustration of this fact is quite
simple and uses the 3-fusion (\ref{rec-3}), see \cite{Roberts} 
for more detail.

\bigskip
\noindent{\bf Lemma (Roberts).} {\it The described above system $C(M,D)$
of meridian and  attaching curves for a handle decomposition $D$ of $M$ gives
a surgery representation of $M\# -M$.}

\bigskip
\noindent This immediately implies the theorem of Turaev and Walker:
\be\label{tw}
{\rm TV}\,(M)=\eta\, I(M\# -M) = |I(M)|^2
\ee
Below we shall see an analog of this relation for a manifold
with boundary. All the facts mentioned make it clear that the TV 
invariant is a natural spin-off of the CS (RTW) invariant.

\bigskip
\noindent{\bf TV inner product.} Recall that the Turaev-Viro inner 
product between the graph states $\mid \Gamma^\psi \rangle$ 
was defined in the previous section as the TV path integral
on $X\times I$. The TV path integral is rigorously defined
by the TV invariant (\ref{tv-triang}). Here we describe how to
compute the inner product in practice.
The prescription we give is from \cite{Turaev}, section 4.d.
We combine it with the chain-mail idea of Roberts \cite{Roberts}
and give this chain-mail prescription.

The product $\langle \Gamma^\psi \mid {\Gamma^\psi}'\rangle$ is
obtained by a certain face model on $X$. Namely, consider
the 3-manifold $X\times I$, where $I$ is the interval $[-1,1]$.
Put $\Gamma^\psi$ on $X \times \{-1\}$ and ${\Gamma^\psi}'$
on $X\times \{1\}$. Both graphs can be projected onto
$X=X\times \{0\}$, keeping track of under- and upper-crossings.
By using an isotopy of $X$ the crossings can be brought into
a generic position of double transversal crossing of edges.
We thus get a graph on $X$, with both 3 and 4-valent vertices.
The 3-valent vertices come from those of $\Gamma^\psi,
{\Gamma^\psi}'$, and 4-valent vertices come from edge
intersections between the two graphs. The inner product
is given by evaluation in $S^3$ of a certain chain-mail
that can be constructed from $\Gamma^\psi,
{\Gamma^\psi}'$. Namely, let us take one 0-framed
link for every face, and one 0-framed link around every edge
of the graph  $\Gamma^\psi\cup{\Gamma^\psi}'$ on $X$.
We get the structure of links at vertices as is shown
in the following drawings:
\be\label{chain-mail}
\lower0.5in\hbox{\epsfig{figure=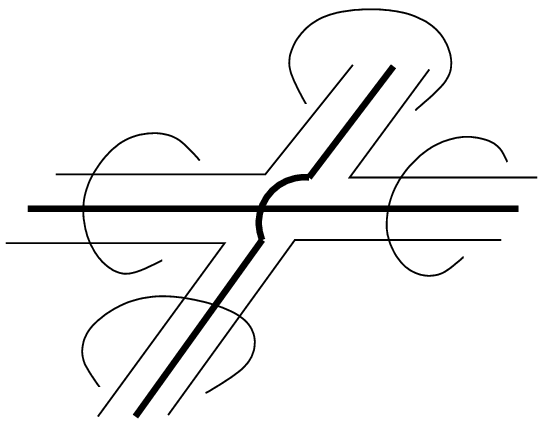,height=1in}}
\hspace{1in}\lower0.35in\hbox{\epsfig{figure=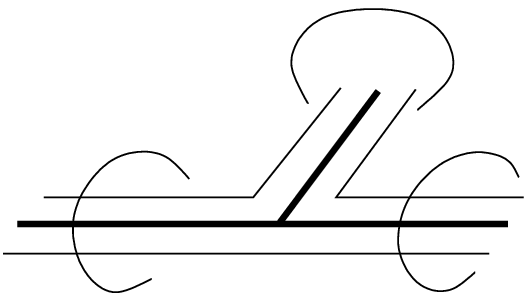,height=0.7in}}
\ee
Denote by $C(\Gamma,\Gamma')$ the obtained collection of
links. The inner product is given by:
\be\label{tv-ip}
\langle \Gamma^\psi \mid {\Gamma^\psi}'\rangle = 
\eta^{V_{\Gamma}+V_{\Gamma'}+V_{int}}
\left( \Gamma^\psi\cup {\Gamma^\psi}' .\,\Omega 
C(\Gamma\cup\Gamma')\right).
\ee
Here $V_\Gamma, V_{\Gamma'}$ are the numbers of 3-valent vertices
of graphs $\Gamma, \Gamma'$ correspondingly, and $V_{int}$ is the
number of 4-valent vertices coming from intersections.
The expression in brackets must be evaluated in $S^3$. Using
the 3-fusion (\ref{rec-3}) one can easily convince oneself
that (\ref{tv-ip}) coincides with the prescription
given in \cite{Turaev}.

We would also like to note an important relation for the
TV inner product that expresses it as the RTW evaluation:
\be\label{tv-ip-evaluation}
\langle \Gamma^\psi \mid {\Gamma^\psi}'\rangle = 
I(X\times S^1, \Gamma^\psi, {\Gamma^\psi}').
\ee
The evaluation is to be carried out in the 3-manifold $X\times S^1$.
This relation that does not seem to have appeared in the literature. 
A justification for it comes from our 
operator/state correspondence, see below. Let us also note that a 
direct proof of a particular
sub-case of (\ref{tv-ip-evaluation}) corresponding to one
of the graphs being zero colored is essentially given by our
proof in the Appendix of the main theorem of section 
\ref{sec:part}. We decided not to attempt a direct proof of 
(\ref{tv-ip-evaluation})
in its full generality. 

\bigskip
\noindent{\bf Turaev theorem.} Let us note
the theorem 7.2.1 from \cite{Turaev-Book}. It states that
the TV invariant for $H$ with the spin network $\Gamma^\psi$
on $X=\partial H$ equals to the RTW evaluation of $\Gamma^\psi$
in $H\cup -H$:
\be\label{theorem}
{\rm TV}\,(H,\Gamma^\psi)=I(H\cup -H, \Gamma^\psi).
\ee
This is an analog of (\ref{tw}) for a manifold with
a single boundary, and is somewhat analogous to our
relation (\ref{tv-ip-evaluation}) for the TV inner product.

\section{Verlinde formula}
\label{sec:ver}

The purpose of this somewhat technical section is to review some facts
about the Verlinde formula for the dimension of the CS Hilbert space. 
Considerations of this section will motivate a more
general formula given in section \ref{sec:part} for the CFT partition function
projected onto a spin network state. This section can be skipped on
the first reading.

\bigskip
\noindent{\bf Dimension of the CS Hilbert space.} Let us first
obtain a formula for the dimension of the CS Hilbert space that explicitly
sums over all different possible states. This can be obtained by
computing the CS
inner product. Indeed, as we have described in section \ref{sec:CS},
a basis in ${\cal H}^{\rm CS}_X$ is given by spin networks
$\Delta^\phi$. With our choice of the normalization of the
3-valent vertices the spin network states $\mid \Delta^\phi \rangle$
are orthogonal but not orthonormal. Below we will show that
the dimension can be computed as:
\be\label{dim-h-cs}
{\rm dim}\,{\cal H}^{\rm CS}_X=
\sum_{\phi} \left( \prod_{{\rm int}\,\,e} {\rm dim}_{\rho_e} \right)
\langle \Delta^\phi \mid \Delta^\phi \rangle =
\sum_{\phi}
\left( \prod_{{\rm int}\,\,e} {\rm dim}_{\rho_e} \right)
I(H\cup -H, \Delta^\phi\cup \Delta^\phi),
\ee
where the sum is taken over the colorings of the internal edges.
The coloring of the edges of $\Delta$ that end at punctures are fixed.

To evaluate $\Delta^\phi\cup{\Delta^\phi}$ we proceed as follows.
Let us project the graph
$\Delta\cup\Delta$ to $X$. We note that there is a canonical
way to do this projection so that there are exactly two
3-valent vertices of $\Delta\cup\Delta$ on each pair of pants,
and there are exactly two edges of $\Delta\cup\Delta$
going through each boundary circle of a pair of pants.
For example, the part of $\Delta\cup\Delta$ projected
on a pair of pants with no punctures looks like:
\be\label{proj-delta}
\lower0.35in\hbox{\epsfig{figure=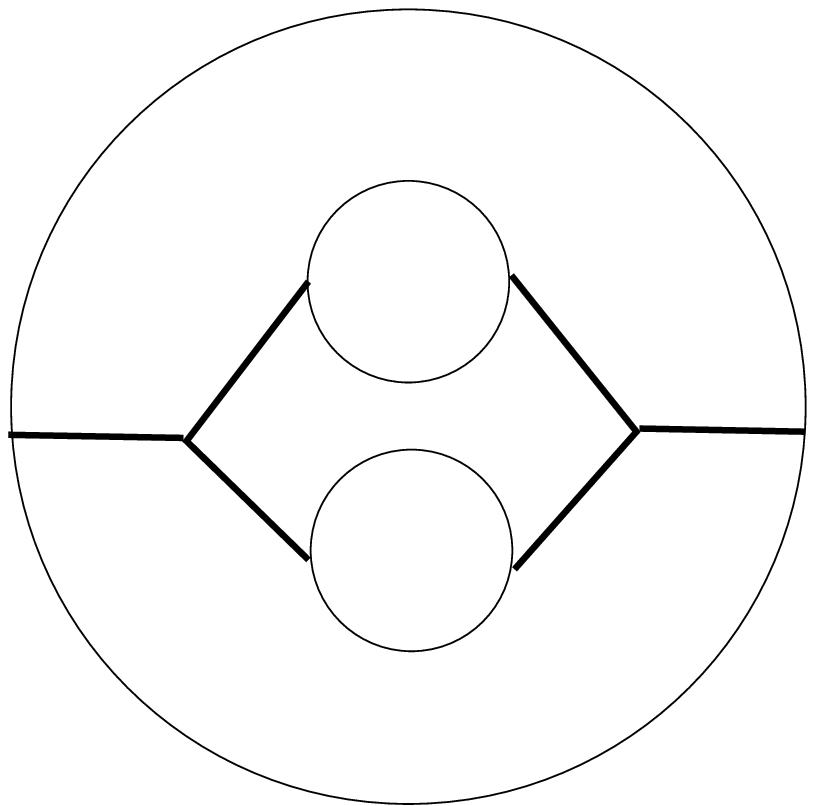,height=0.7in}}
\ee
One gets a similar structure when projecting on a pair
of pants with punctures. In that case the two holes in the
center are replaced by punctures and loose edges of
$\Delta$ are connected  at the punctures to the loose edges of
the other copy of $\Delta$.

Let us now form a link $L_\Delta$ whose components are
circles along which one glues the pant boundaries together.
There are $3g+n-3$ such circles, in one-to-one
correspondence with internal edges of $\Delta$. We push
all components of $L_\Delta$ slightly out of $X$. Using the
prescription of the Appendix of \cite{Turaev} for computing 
the RTW evaluation of $M$ with a graph inserted, one obtains:
\be\label{ver-general}
I(H\cup -H, \Delta^\phi\cup \Delta^\phi) = \eta^{3g+n-3} 
\left( \Delta^\phi\cup\Delta^\phi . \Omega L_\Delta\right).
\ee
The evaluation on the right hand side is to be taken in $S^3$.
This relation establishes (\ref{dim-h-cs}). 
Indeed, there are exactly two edges of $\Delta\cup\Delta$
linked by every component of $L_\Delta$. Using the
2-fusion we get them connected at each pair of pants, times
the factor of $\eta^{-1}/{\rm dim}_{\rho_e}$.
The factors of $\eta$ are canceled
by the pre-factor in (\ref{ver-general}), and
the factors of $1/{\rm dim}_{\rho_e}$ are canceled by the
product of dimensions in (\ref{dim-h-cs}). What remains is the sum over
the colorings of the internal edges of the product of $N_{ijk}$ for
every pair of pants. This gives the
dimension. This argument also shows that the states $\mid \Delta^\phi\rangle$
with different coloring $\phi$ are orthogonal.

\bigskip
\noindent{\bf Computing the dimension: Verlinde formula.} The sum over
colorings of the internal edges in (\ref{ver-general}) can be computed.
This gives the Verlinde formula. Let us sketch a simple proof of it,
for further reference.

We first observe that, using the 3-fusion, the Verlinde formula
for the 3-punctured sphere can be obtained as a chain-mail. Namely,
\be\label{obs}
\eta^{-1} N_{ijk} =
\lower0.5in\hbox{\epsfig{figure=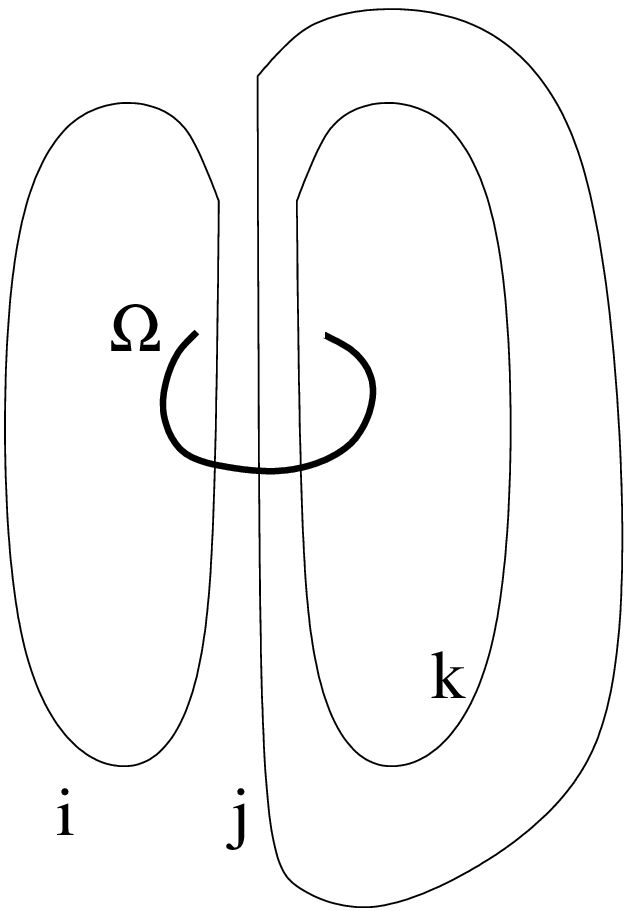,height=1in}}
\ee
The Verlinde formula for $N_{ijk}$ can be obtained by using the
definition (\ref{omega}) of $\Omega$ and the recoupling identity
(\ref{rec-s}) of the Appendix. The computation is as follows:
\be
\eta^{-1} N_{ijk} =
\eta \sum_l {\rm dim}_l \,\,
\lower0.5in\hbox{\epsfig{figure=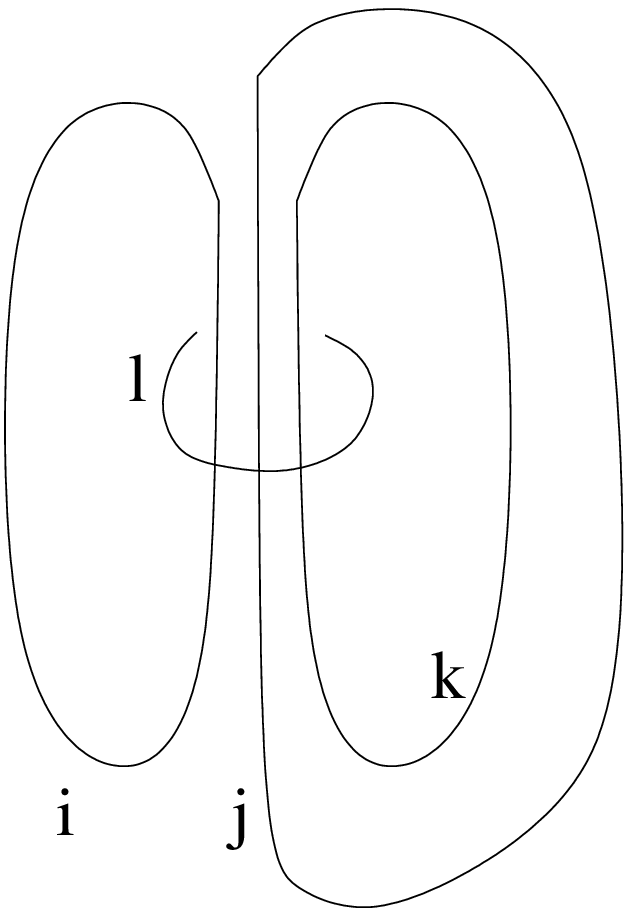,height=1in}}=
\sum_l S_{il} \,\,
\lower0.5in\hbox{\epsfig{figure=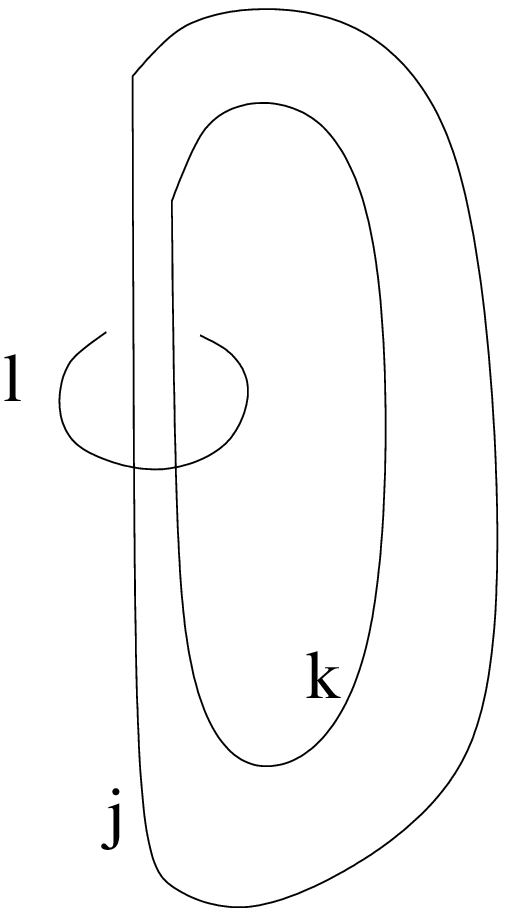,height=1in}}=
\sum_l {S_{il} S_{jl}\over \eta\,{\rm dim}_l} \,\,
\lower0.4in\hbox{\epsfig{figure=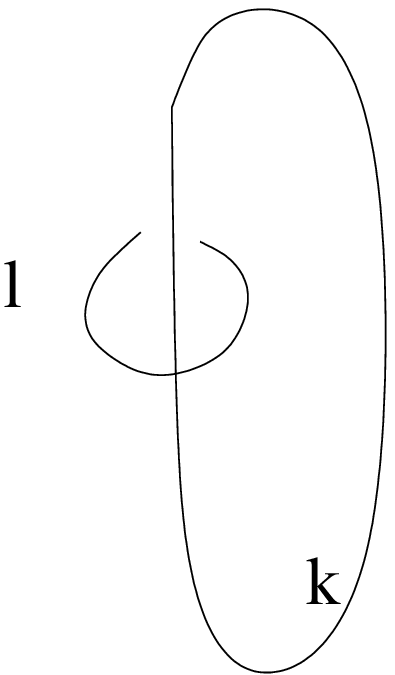,height=0.8in}}=
\eta^{-1} \sum_l {S_{il} S_{jl} S_{kl} \over S_{0l}}.
\ee
This is the Verlinde formula (\ref{ver-formula}) for the
case of a 3-punctured sphere. We have used the fact
that $\eta\, {\rm dim}_l =S_{0l}$. The above proof of
the Verlinde formula for $N_{ijk}$ is essentially that
from \cite{Witten-Jones}.

The general Verlinde formula (\ref{ver-formula}) can be obtained using
a pant decomposition of $X$ and taking a sum over labelings of the
internal edges of $\Delta$ of the product of $N_{ijk}$ one for
every pair of pants. To get (\ref{ver-formula}) one just has
to use the unitarity $\sum_l S_{il} S_{jl}=\delta_{ij}$ of the
S-matrix. 

\bigskip
\noindent{\bf Verlinde formula using graph $\Gamma$: no punctures.}
Here we find a different representation of the Verlinde dimension.
It was noticed in \cite{Boulatov} that the Verlinde formula
can be obtained using a certain
gauge theory on a graph on $X$. Here we re-interpret this result
using a chain-mail. We first derive a formula for a Riemann surface
without punctures. It is obtained by starting from a graph $\Gamma$
corresponding to a surface with some number $n$ of punctures.
Then a sum is carried over the labels at the punctures, so the end
result depends only on the genus $g$, but not on $n$.

Consider a fat tri-valent graph $\Gamma$
that represents $X_{g,n}$. Let us form a chain-mail
$C(\Gamma)$ as follows. Let us 
introduce a curve for every face of the fat graph $\Gamma$,
and a linking curve around every of
$3(2g+n-2)$ edges of $\Gamma$, so that the obtained structure of
curves at each 3-valent vertex is as in (\ref{chain-mail}). Insert the
element $\Omega$ along each component of $C(\Gamma)$, and evaluate the
result in $S^3$. What is evaluated is just the chain-mail for
$\Gamma$, no spin network corresponding to $\Gamma$ is inserted. 
We get the following result:

\bigskip
\noindent{\bf Theorem (Boulatov)} {\it
The dimension of the Hilbert space of CS states
on $X_g$ is equal:}
\be\label{ver-compact}
{\rm dim}\,{\cal H}^{\rm CS}_{X_g} = \eta^{V_\Gamma}\,
\Omega C(\Gamma).
\ee
{\it The expression on the right hand side is independent of the graph
$\Gamma$ that is used to evaluate it.}
\bigskip

To prove this formula we use the 2-strand fusion.
We get that all of the $n$ different colorings on the links of
$\Gamma$ become the same. Denote by $\rho$ the corresponding
representation. The result is then obtained by a simple counting.
Each of $3(2g+n-2)$ of links around edges introduces the factor
of $\eta^{-1}/{\rm dim}_\rho$. Every of $2(2g+n-2)$ vertices
of $\Gamma$ gives a factor of ${\rm dim}_\rho$. Each of $n$
faces of $\Gamma$ gives another factor of $\eta {\rm dim}_\rho$.
All this combines, together with the pre-factor to give:
\be
{\rm dim}\,{\cal H}^{\rm CS}_{X_g} = \sum_\rho (\eta\, {\rm dim}_\rho)^{2-2g},
\ee
which is the Verlinde formula (\ref{ver-formula}) for the
case with no punctures.

\section{Operator/state correspondence}
\label{sec:rel}

This section is central to the paper. Here we discuss a one-to-one
correspondence between observables of CS theory and quantum
states of TV theory. The fact that the algebra of observables in
CS theory is given by graphs is due to \cite{Fock,Alekseev},
see also references below. The notion of the connecting 3-manifold
$\tilde{M}$ is from \cite{Schweigert,Gawedzki}. The
operator/state correspondence of this section, as well as the
arising relation between the CS and TV Hilbert spaces,
although to some extent obvious, seem new.

\bigskip
\noindent{\bf CS observables and relation between the Hilbert spaces.} 
We have seen that a convenient
parameterization of the moduli space ${\cal A}/{\cal G}$ is
given by the graph $\Gamma$ connections.
An expression for the CS Poisson structure in terms of graph connections
was found in \cite{Fock}. A quantization of the corresponding
algebra of observables was developed in \cite{Alekseev,AM,AS,AGS-1,AGS-2},
see also \cite{BNR} for a review. As we have seen in section \ref{sec:TV}
a complete set of functionals on ${\cal A}/{\cal G}$ is
given by spin networks. Spin networks thus become operators
$\hat{\Gamma}^\psi$ in the CS Hilbert space ${\cal H}^{\rm CS}_X$.
We therefore get a version of an operator/state correspondence,
in which TV states correspond to observables of CS theory.

The fact that a CS/TV operator/state correspondence must
hold follows from the relation between the phase spaces of the
two theories. Namely,
as we have seen in section \ref{sec:TV}, the TV phase space
is given by two copies of the phase space of Chern-Simons theory:
${\cal P}^{\rm TV}= {\cal P}^{\rm CS}\otimes \bar{{\cal P}}^{\rm CS}$, 
where the two copies have opposite Poisson structures. This 
means that in the quantum theory the following relation must hold:
\be\label{isom}
{\cal H}^{\rm TV}_X \sim {\cal H}^{\rm CS}_X\otimes
{\cal H}^{\rm CS}_{-X}\sim {\rm End}\,\left({\cal H}^{\rm CS}_X\right).
\ee
Thus, the TV Hilbert space is isomorphic to the direct
product of two copies of ${\cal H}^{\rm CS}$. The above isomorphism,
which we shall denote by $\bf I$,
identifies the TV spin network states $\mid \Gamma^\psi\rangle$ 
with the CS spin network observables $\hat{\Gamma}^\psi$. This
statement deserves some explanation. The TV spin network states
are wave functionals of the connection 
$\ul{\bf w}: \Phi(\ul{\bf w})=\Phi(A_z+B_z,A_{\bar{z}}+
B_{\bar{z}})$, whereas
Chern-Simons states are functionals $\Psi(A_z,B_{\bar z})$.
Thus, the isomorphism (\ref{isom}) can be understood as a change of 
polarization. Being a change of polarization it 
intertwines the operator algebras acting on the two sides of
(\ref{isom}). The polarisation we have
choosen for the TV Viro model is the one for which 
$\hat{\bf e}\sim(\hat{A}-\hat{B})$ acts trivially on the TV vaccum state.
Using the intertwinning property of $\bf I$ this means that
${\bf I}(\mid0\rangle_{\rm{TV}})$ is commuting with all 
CS operators $\hat{\Gamma}^\psi$. It is therefore
proportional to the identity in ${\rm End}\,\left({\cal H}^{\rm CS}_X\right)$.
It follows from here that the operator that corresponds to
the TV state $\mid \Gamma^\psi \rangle$ is the CS spin network operator:
\be
{\bf I}(|\Gamma^\psi \rangle_{\rm{TV}})=\hat{\Gamma}^\psi
{\bf I}(|0\rangle_{\rm{TV}})\propto \hat{\Gamma}^{\psi}.
\ee
Thus, the described isomorphism (\ref{isom}) given by the change
of polarization indeed identifies TV graph states with the CS
spin network operators.

Another important fact is as follows. Being a change of polarization,
the isomorphism (\ref{isom}) preserves the inner product. Since the
inner product on the right hand side of (\ref{isom}) is just the
CS trace, we get an important relation:
\be\label{tr-prod}
{\rm Tr}_{\rm CS} \left( \hat{\Gamma} \hat{\Gamma'}
\right) = \langle \Gamma \mid {\Gamma}' \rangle_{\rm TV}.
\ee
In other words, the trace of the product of operators in the CS Hilbert space
is the same as the inner product in the TV theory. This relation
is central to the operator/state correspondence under consideration.
Let us now describe the isomorphism (\ref{isom}) more explicitly.

\begin{figure}
\centerline{\hbox{\epsfig{figure=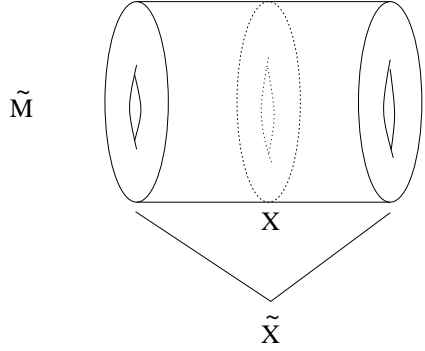,height=1.8in}}}
\caption{The manifold $\tilde{M}$.}
\label{fig:double}
\end{figure}

\bigskip
\noindent{\bf Connecting manifold $\tilde{M}$.} A very effective 
description of the above operator/state correspondence
uses the ``connecting manifold'' $\tilde{M}$. It is a 3-manifold whose boundary
is the Schottky double $\tilde{X}$ of the Riemann surface $X$.
Recall that the Schottky double of a Riemann surface $X$ is another
Riemann surface $\tilde{X}$. For the case of a closed $X$,
the surface $\tilde{X}$ consists of two disconnected
copies of $X$, with all moduli replaced by their complex
conjugates in the second copy. For $X$ with a boundary
(the case not considered in this paper, but of relevance
to the subject of boundary CFT, see, e.g., \cite{Schweigert,Gawedzki})
the double $\tilde{X}$ is obtained by taking two copies of $X$
and gluing them along the boundary. Consider a 3-manifold
\be
\tilde{M}=\tilde{X}\times [0,1] /\sigma,
\ee
where $\sigma$ is an anti-holomorphic map such that
$\tilde{X}/\sigma=X$, and $\sigma$ reverses the ``time'' direction.
See, e.g., \cite{Schweigert} for more detail on the construction
of $\tilde{M}$. The manifold $\tilde{M}$ has
a boundary $\partial \tilde{M}=\tilde{X}$, and the original
surface $X$ is embedded into $\tilde{M}$. For the case of
a closed $X$, relevant for this paper, the manifold
$\tilde{M}$ has the topology $X\times I$, where
$I$ is the interval $I=[0,1]$, see Fig.~\ref{fig:double}.

Consider the RTW evaluation
of a spin network $\Gamma^\psi$ in $\tilde{M}$. It gives
a particular state in ${\cal H}^{\rm CS}_{\tilde{X}}$:
\be\label{state-op}
I(\tilde{M},\Gamma^\psi) \in {\cal H}^{\rm CS}_{\tilde{X}}.
\ee
However, we have:
\be
{\cal H}^{\rm CS}_{\tilde{X}} \sim {\cal H}^{\rm CS}_X\otimes
{\cal H}^{\rm CS}_{-X}\sim {\rm End}\left({\cal H}^{\rm CS}_X\right).
\ee
Thus (\ref{state-op}) gives an operator in ${\cal H}^{\rm CS}_X$
for every graph state $\mid \Gamma^\psi \rangle\in {\cal H}^{TV}_X$.

\bigskip
\noindent{\bf Operator product.} In the realization described the product
of two operators $\hat{\Gamma}^\psi, \hat{\Gamma}^\psi{}'$ is an
element of ${\cal H}^{\rm CS}_{\tilde{X}}$ obtained by evaluating
in $\tilde{M}$ both $\Gamma^\psi$ and ${\Gamma^\psi}'$:
\be
I(\tilde{M},\Gamma^\psi,{\Gamma^\psi}' ) \in {\cal H}^{\rm CS}_{\tilde{X}}.
\ee

\bigskip
\noindent{\bf Trace.} The trace of an operator $\hat{\Gamma}^\psi$
is obtained by gluing the two boundaries of $\tilde{M}$
to form a closed manifold of the topology $X \times S^1$:
\be
{\rm Tr}_{\rm CS}\,
\left( \hat{\Gamma}^\psi \right) = I(X\times S^1, \Gamma^\psi).
\ee
One can similarly obtain the trace of an operator product:
\be
{\rm Tr}_{\rm CS}\left(\hat{\Gamma}^\psi
\hat{\Gamma}^\psi{}'\right)=I(X\times S^1, \Gamma^\psi, \Gamma^\psi{}').
\ee
In view of (\ref{tr-prod}), the above relation establishes 
(\ref{tv-ip-evaluation}).

\bigskip
\noindent{\bf Identity operator.} It is easy to see that the
operator/state correspondence defined by (\ref{state-op}) 
is such that the zero colored graph $\Gamma^0$ corresponds
to the identity operator in the CS Hilbert space:
\be\label{zero-identity}
\hat{\Gamma}^0 = \hat{I}.
\ee
Indeed, insertion of $\Gamma^0$ into $\tilde{M}$ is same
as $\tilde{M}$ with no insertion, whose RTW evaluation
gives the identity operator in ${\cal H}^{\rm CS}$.

\bigskip
\noindent{\bf Matrix elements.} We recall 
that a basis in ${\cal H}^{\rm CS}_X$ 
is obtained by choosing a pant decomposition of $X$,
or, equivalently, choosing a tri-valent graph $\Delta$,
with a coloring $\phi$. The matrix elements
$\langle \Delta^\phi \mid \hat{\Gamma}^\psi \mid {\Delta^\phi}'\rangle$
are obtained by the following procedure. Take a handlebody
$H$ with a graph $\Delta^\phi$ in it, its loose ends ending
at the punctures. The boundary of $H$ is $X$, so that
we can glue $H$ from the left to $\tilde{M}$. One similarly
takes $-H$ with ${\Delta^\phi}'$ in it, and glues it
to $\tilde{M}$ from the right. One connects the punctures on
the boundary of $H$ to those on the boundary of $-H$ by
strands inside $\tilde{M}$. What one gets is a closed
manifold of the topology $H\cup -H$, with closed graphs
$\Delta^\phi\cup{\Delta^\phi}'$ and $\Gamma^\psi$ sitting
inside it. The matrix elements are obtained as the
evaluation:
\be\label{relation}
\langle \Delta^\phi \mid \hat{\Gamma}^\psi \mid {\Delta^\phi}'\rangle=
I(H\cup -H,\Delta^\phi\cup{\Delta^\phi}',\Gamma^\psi).
\ee

\section{CFT partition function as a state}
\label{sec:part}

Here we interpret the CFT partition function (correlator) as
a particular state in the Hilbert space of TV theory. We also
compute components of this state in the basis of states
given by spin networks.

\bigskip
\noindent{\bf CFT partition function.} The partition function of
any CFT holomorphically factorizes. To understand this holomorphic
factorization, and the relation to the chiral TQFT, it is most
instructive to consider the partition function as a function of
an external connection. Namely, let CFT be the WZW model coupled to
an external connection (gauged model), and consider
its partition function
$Z^{\rm CFT}_X[{\bf m},\bar{\bf m},{\bf z},\bar{\bf z},
\ul{A}_z,\ul{A}_{\bar{z}}]$ on $X$. Note that
no integration is carried over $\ul{A}$ yet.
Thus, the above quantity is not what is usually called the
gauged WZW partition function. The later is obtained by integrating
over $\ul{A}$. The introduced partition function
depends on the moduli (both holomorphic
and anti-holomorphic) ${\bf m}, \bar{\bf m}$, on positions of
insertions of vertex operators coordinatized by ${\bf z},\bar{\bf z}$,
and on both the holomorphic and anti-holomorphic components of the
connection $\ul{A}$ on $X$. The partition function
holomorphically factorizes according to:
\be\label{z-gauged}
Z^{\rm CFT}_X[{\bf m},\bar{\bf m},{\bf z},\bar{\bf z},
\ul{A}_z,\ul{A}_{\bar{z}}]=
\sum_i \Psi_i[{\bf m},{\bf z},\ul{A}_z]
\bar{\Psi}_i[\bar{\bf m},\bar{\bf z},\ul{A}_{\bar{z}}].
\ee
Here $\Psi_i[{\bf m},{\bf z},\ul{A}_z]$ are the (holomorphic) conformal
blocks, which can be thought of as forming a basis in the
Hilbert space ${\cal H}^{\rm CS}_X$ of CS theory on $X$.
More precisely, there is a fiber bundle over the moduli space
${\cal M}_{g,n}$ of Riemann surfaces of type $(g,n)$ with
fibers isomorphic to ${\cal H}^{\rm CS}_{X_{g,n}}$. The
conformal blocks are (particular) sections of this bundle,
see \cite{FS} for more detail. Note that the sum in (\ref{z-gauged}) is
finite as we consider a rational CFT. As was explained in
\cite{Witten-Holomorphic}, the usual CFT partition function
is obtained by evaluating (\ref{z-gauged}) on the ``zero''
connection. The formula (\ref{z-gauged}) then gives the factorization
of the usual partition function, with $\Psi_i[{\bf m},{\bf z},0]$
being what is usually called the Virasoro conformal blocks.

Instead of evaluating (\ref{z-gauged}) on the zero connection
one can integrate over $\ul{A}$. The result is the partition function
of the gauged model, which gives the dimension of the CS
Hilbert space:
\be
{\rm dim}\,{\cal H}^{\rm CS}_X = {1\over {\rm Vol}\,{\cal G}}
\int_{\cal A} {\cal D}\ul{A}\,\,
Z^{\rm CFT}_X[{\bf m},\bar{\bf m},{\bf z},\bar{\bf z},
\ul{A}_z,\ul{A}_{\bar{z}}].
\ee
The value of the integral on the right hand side is independent
of moduli (or positions of insertion points).

A particular basis of states in ${\cal H}^{\rm CS}_X$
was described in section \ref{sec:CS} and is given by
states $\mid \Delta^\phi \rangle$. Let us use these states
in the holomorphic factorization formula (\ref{z-gauged}).
We can therefore think of the partition function (correlator)
as an operator in the CS Hilbert space:
\be\label{part-state}
\hat{Z}^{\rm CFT}_X = \sum_\phi
\left( \prod_{{\rm int}\,\,e} {\rm dim}_{\rho_e} \right)
\mid \Delta^\phi \rangle \otimes
\langle \Delta^\phi \mid.
\ee
The dimension of the CS Hilbert space is obtained by taking the CS
trace of the above operator, which gives (\ref{dim-h-cs}).

The CFT partition function (\ref{part-state}) 
is the simplest possible modular invariant (the diagonal) that can
be constructed out of the chiral CFT data. There are other possible
modular invariants, and it is an ongoing effort to try to
understand and classify different possibilities, see, e.g.,
the recent paper \cite{Schw-1}. In this paper we only consider
and give a TV interpretation of the simplest invariant 
(\ref{part-state}). Our TV interpretation might prove useful also
for the classification program, but we do not pursue this.

\bigskip
\noindent{\bf CFT Partition function as a state.} The
formula (\ref{part-state}) for the partition function,
together with the operator/state correspondence of the
previous section imply that $Z^{\rm CFT}_X$
can be interpreted as a particular state in the TV Hilbert space.
We introduce a special notation for this state:
\be\label{state}
\mid Z^{\rm CFT}_X \rangle \in {\cal H}^{\rm TV}_X.
\ee
In order to characterize this state we first of all note that
$\hat{Z}^{\rm CFT}_X$ is just the identity operator in
${\cal H}^{\rm CS}_X$:
\be
\hat{Z}^{\rm CFT}_X = \hat{I}.
\ee
The representation (\ref{part-state}) 
gives the decomposition of the identity over a complete
basis of states in ${\cal H}^{\rm CS}_X$. Using (\ref{zero-identity})
we see that the state $\mid Z^{\rm CFT}_X \rangle$
is nothing else but the spin network state with zero coloring,
together with a set of strands labelled with representations
$\bf R$ and taking into account the punctures:
\be\label{z-zero}
\mid Z^{\rm CFT}_X \rangle = \mid \Gamma^0, {\bf R} \rangle.
\ee

Another thing that we are interested in is the components of
$\mid Z^{\rm CFT}_X \rangle$ in the basis of 
spin networks $\mid \Gamma^\psi \rangle$. In view of 
(\ref{tv-ip-evaluation}) we have:
\be\label{part-spin-net}
\langle \Gamma^\psi \mid Z^{\rm CFT}_X \rangle =
I(X\times S^1, {\bf R}, \Gamma^\psi).
\ee
The evaluation in $X\times S^1$ is taken in the presence of
$n$ links labelled by representations $\bf R$.
Note that all the dependence on the moduli of $X$ is lost
in (\ref{part-spin-net}). However, the coloring $\psi$ of $\Gamma$
can be thought of as specifying the ``geometry'' of $X$, see more
on this below.

\bigskip
\noindent{\bf Zero colored punctures.} Here, to motivate the
general formula to be obtained below, we deduce an expression
for (\ref{part-spin-net}) for the case where the colors at
all punctures are zero. In this case there is no extra links
to be inserted in $X\times S^1$, and (\ref{part-spin-net})
reduces to $\langle \Gamma^\psi \mid \Gamma^0 \rangle$.
This can be evaluated using the prescription (\ref{tv-ip}).
One immediately obtains:
\be\label{part-compact}
\langle \Gamma^\psi \mid \Gamma^0 \rangle=
\eta^{V_\Gamma}
\left(\Gamma^\psi .\Omega C(\Gamma) \right) =
\eta^{2-2g} \sum_{\{\rho_f\}} \prod_{f\in F_\Gamma}
{\rm dim}_{\rho_f} \prod_{v\in V_\Gamma} (6j)_v.
\ee
Here $C(\Gamma)$ is the chain-mail for $\Gamma$, as defined
in the formulation of the theorem (\ref{ver-compact}). In the last formula
the sum is taken over irreducible representations labelling
the faces of the fat graph $\Gamma$, the product of 6j-symbols is
taken over all vertices of $\Gamma$, and the 6j-symbols
$(6j)_v$ are constructed out of three representations labelling the
edges incident at $v$, and three representations labelling
the faces adjacent at $v$. The last formula is obtained using
the 3-fusion recoupling identity (\ref{rec-3}).

\bigskip
\noindent{\bf Verlinde formula.} The
dimension of the CS Hilbert space can be obtained as the
inner product of $\mid Z^{\rm CFT}_X \rangle$ with the
``vacuum'' state $\mid \Gamma^0 \rangle\in{\cal H}^{\rm TV}_X$,
which corresponds to the spin network with zero (trivial
representation) coloring on all edges:
\be\label{ver}
{\rm dim}\,{\cal H}^{\rm CS}_X = \langle \Gamma^0 \mid Z^{\rm CFT}_X \rangle.
\ee
The expression (\ref{ver}) gives
an unusual perspective on the Verlinde formula:
it appears as a particular case of a more general
object (\ref{part-spin-net}).

\bigskip
\noindent{\bf General formula.} Here we find the 
result of the evaluation (\ref{part-spin-net}). As we have just explained,
(\ref{part-spin-net}) must reduce to the Verlinde formula (\ref{ver-formula})
when the graph $\Gamma$ has zero colors. We have seen in 
section \ref{sec:ver} that, at least for the case with
no punctures, the Verlinde formula can
be obtained from the chain-mail $C(\Gamma)$ with no graph
$\Gamma$ inserted. We have also seen in (\ref{part-compact}) that
for the case with no punctures the quantity (\ref{part-spin-net})
is given by the evaluation of $C(\Gamma)$ together with the
graph. Thus, a natural proposal for (\ref{part-spin-net}) is
that it is given by the evaluation (\ref{ver-compact}), with the
graph $\Gamma$ added, and with an additional
set of curves taking into account the punctures. This results in:

\bigskip
\noindent{\bf Main Theorem.} {\it The CFT partition function (correlator) 
projected onto a spin network state is given by:}
\be\label{main}
\langle \Gamma^\psi \mid Z^{\rm CFT}_X \rangle = \eta^{2-2g-n}
\sum_{\{\rho_f\}} \prod_{f\in F_\Gamma} S_{\rho_i \rho_{f_i}}
\prod_{v\in V_\Gamma} (6j)_v.
\ee

\bigskip
\noindent A proof is given in the Appendix.

\section{Discussion}
\label{sec:dis}

Thus, the CFT partition function (correlator) receives the
interpretation of a state of TV theory. This state is the
TV ``vacuum'' given (\ref{z-zero}) by the graph with zero 
coloring. Thus, quite a non-trivial object from the point of
view of the CFT, the partition function 
receives a rather simple interpretation in the TV theory.

We note that, apart from the partition function state 
$\mid Z_{\rm CFT}\rangle$, there is another state in
${\cal H}^{\rm TV}$ with a simple CS interpretation.
This is the state that can be denoted as 
\be
\mid H\rangle \in {\cal H}^{\rm TV}.
\ee
It arises as the TV partition function for a handlebody $H$.
The TV invariant (\ref{tv-triang}) for a manifold with boundary 
has the interpretation of the TV inner product of $\mid H\rangle$
with a spin network state:
\be
TV(H, \Gamma^\psi) = \langle H \mid \Gamma^\psi \rangle.
\ee
In view of the Turaev theorem (\ref{theorem})
\be
\langle H \mid \Gamma^\psi \rangle = I(H\cup -H, \Gamma^\psi).
\ee
From this, and the relation (\ref{relation}) for the
matrix elements it can be seen that the state $\mid H \rangle$
corresponds in CS theory to the operator 
\be\label{proj}
\hat{H} = \mid \Delta^0 \rangle \otimes \langle \Delta^0 \mid,
\ee
which is just the projector on the CS ``vacuum'' state
$\Delta^0$, given by the zero colored pant decomposition
graph $\Delta$. We note that the TV state $\mid H \rangle$ has a rather
non-trivial expression when decomposed into the spin network
basis. Thus, the described relation between CS and TV theories
(the operator/state correspondence) is a non-trivial
duality in that simple objects on one side correspond to
non-trivial objects on the other: CFT correlators, non-trivial
from the point of view of CS, are the TV ``vacuum'' states;
the non-trivial TV handlebody state $\mid H\rangle$ is a
rather trivial ``vacuum'' projector on the CS side. 

We would like to emphasize that the CFT partition function
state $\mid Z^{\rm CFT}_X\rangle$ does not coincide 
with the TV partition function state $\mid H\rangle$ 
on a handlebody $H$. Thus, we can only interpret 
the CFT partition function as the TV vacuum (\ref{z-zero}).
It does not seem to arise as a TV partition function
corresponding to some 3-manifold $M$. Thus, the CFT/TQFT 
holographic correspondence that we are discussing is
rather subtle in that CFT partition function is a
state in the boundary TV Hilbert space, but it is not
a HH state arising as the path integral over some $M$
that has $X$ as the boundary.

Thus, we have seen that there are two TV states that correspond
to CFT modular invariants: one is the TV vacuum (\ref{z-zero})
that gives the diagonal modular invariant, the other is
the handlebody state $\mid H\rangle$ that gives the trivial
modular invariant (\ref{proj}). An interesting question is
what other states in TV give CFT modular invariants. An answer
to this question may be instrumental in understanding the
structure of rational CFT's, see the recent paper \cite{Schw-1}
for a discussion along these lines.  

Let us now discuss a physical interpretation of the formula
(\ref{main}). We note that the object (\ref{part-spin-net})
can be interpreted as the CFT partition function on a surface
$X$ whose ``geometry'' is specified by the state $\mid \Gamma^\psi \rangle$.
This ``geometry'' should not be confused with the conformal geometry
of $X$, on which the usual CFT partition function depends. Once
the state $\mid Z^{\rm CFT}_X \rangle$ is projected onto
$\mid \Gamma^\psi \rangle$ the dependence on the moduli of
$X$ is traded for the dependence on the coloring $\psi$ of
$\Gamma$. All the dependence on the moduli is encoded in the
spin network states. Let us first discuss the dependence on the
``geometry'' as specified by the colored graph $\Gamma^\psi$,
and then make comments as to the dependence of $\mid \Gamma^\psi \rangle$
on the moduli.

To understand the spin network $\Gamma^\psi$ as specifying the
``geometry'' of $X$ we recall, see section \ref{sec:TV},
that $\mid \Gamma^\psi \rangle$ are eigenstates of the
``momentum'' operators ${\bf e}\sim \partial/\partial \ul{\bf w}$.
In this sense they are states of particular configuration
of the ${\bf e}$ field on the boundary. To understand this
in more detail let us consider the TV partition function
$TV(H, \Gamma^\psi)$. Let us take the simple example of 
the 4-punctured sphere. Thus, we take $H=B^3$, a 3-ball. We will 
put all representations at the punctures to be trivial. In view of the
Turaev theorem (\ref{theorem}) $TV(B^3, \Gamma^\psi)=I(S^3, \Gamma^\psi)$.
Thus, for $X=S^2$, the TV invariant is given simply 
by the evaluation of the spin network
$\Gamma^\psi$ in $S^3$. In our simple example of the
4-punctures sphere this evaluation is a single 6j-symbol.
Let us now restrict ourselves to the
case $G={\rm SU}(2)$. As we have mentioned above, the
TV theory in this case is nothing else but 3d gravity with
positive cosmological constant. On the other hand, it is
known that the quantum (6j)-symbol has, for large $k$ and
large spins, an asymptotic of the exponential of the
classical Einstein-Hilbert action evaluated inside the
tetrahedron:
\be\label{6j}
(6j) \sim e^{i S_{TV}[tet]} + {\rm c.\,\, c.}
\ee
This fact was first observed \cite{PR} by Ponzano and Regge for the
classical (6j)-symbol. In that case one evaluates the classical
gravity action inside a flat tetrahedron. The action reduces to a
boundary term (the usual integral of the trace of the extrinsic curvature
term), which for a tetrahedron is given by the so-called Regge action:
\be\label{action-flat}
S_{TV}[tet,\Lambda=0]\sim \sum_e l_e \theta_e,
\ee
where the sum is taken over the edges of the tetrahedron, and
$l_e, \theta_e$ are the edge length and the dihedral angle
at the edge correspondingly. Dihedral angles are fixed once all
the edge length are specified. Ponzano and Regge observed that
the (6j)-symbol has the asymptotic of (\ref{6j}) with
the action given by (\ref{action-flat}) if spins labelling
the edges are interpreted as the length
of edges. A similar (\ref{6j}) interpretation is true for the 
${\rm SU}_q(2)$ (6j)-symbol, as was shown in \cite{MT}.
The gravity action in this case is that with a positive
cosmological constant $\Lambda=(k/2\pi)^2$, and is evaluated
in the interior of tetrahedron in $S^3$ whose edge length are given 
by spins. To summarize, in these examples 
the (6j)-symbol gets the interpretation
of the exponential of the classical gravity action evaluated
inside a tetrahedron embedded in either $\R^3$ or $S^3$,
depending on whether one takes the classical limit $k\to\infty$ or
considers a quantum group with finite $k$. The tetrahedron itself
is fixed once all edge length are specified. The edge length
are essentially given by the spins. We also note that
the graph $\Gamma$ in this example is the dual graph to the
triangulated boundary of the tetrahedron in question.

Thus, the TV partition function
(given by a single (6j)-symbol) inside a 4-punctured sphere
(tetrahedron) has the interpretation of the gravity
partition function inside the tetrahedron with its boundary
geometry (edge length) fixed by the spins. 
This interpretation of $\Gamma^\psi$ is valid also for other
surfaces. One should think of $\Gamma^\psi$ as specifying the
geometry ${\bf e}$ on $X$. The TV invariant is, in the
semi-classical limit of large representations, dominated
by the exponential of the classical action evaluated inside
the handlebody. The geometry inside is completely determined
by the geometry of the surface, in other words, the spins.
The interpretation is valid not only for ${\rm SU}(2)$, but
also for other groups. In such a general case the notion
of ``geometry'' is more complicated, as described
by the field $\bf e$ and the TV action (\ref{tv-action}).

The bottom line is that the TV spin network states 
$\mid \Gamma^\psi\rangle$ should
be thought of as specifying the ``geometry'' of $X$. The quantity
(\ref{main}) then receives the interpretation of the CFT
partition function on a surface $X$ whose ``geometry''
is specified by $\Gamma^\psi$.

The other question is how the states $\mid \Gamma^\psi \rangle$
depend on the moduli of the surface. The fact that the
graph $\Gamma$ is the same as the one used in the Penner
\cite{Penner} coordinatization of the moduli space suggests
that this dependence may be not very complicated. In fact,
we believe that for the groups $\SL(2,\R)$ or $\SL(2,\C)$ that
are relevant in the description of the moduli spaces,
the dependence is rather simple: the described above
``geometry'' in this case must coincide with the usual
conformal geometry of the surface. An argument for this is as
follows. In the Penner coordinatization
of the moduli space, or in any of its versions \cite{Fock-1,Kashaev}
the moduli are given by prescribing a set of real numbers: one for
each edge of the graph $\Gamma$. The numbers specify how two
ideal triangles are glued together across the edge,
see \cite{Fock-1,Kashaev} for more detail. For the case when
$G=\SL_q(2,\R)$, as is relevant for, e.g., Liouville theory,
see \cite{Teschner}, the representations are also labelled
by a single real number. We believe that the Penner coordinates
and the representations that label the edges are simply dual
to each other, in the sense of duality between the
conjugacy classes of elements in the group and its irreducible
representations. A similar proposal for the relation between
the ${\rm SL}(2)$ spin and length was made in \cite{Verlinde-H}.
Thus, there is some hope that the dependence
$\mid \Gamma^\psi \rangle$ on the moduli can be understood
rather explicitly, at least for some groups. Having this
said we note that considerations of the present paper
do not immediately generalize to the case of non-compact groups,
relevant for the description of the moduli spaces.
It is an outstanding problem to develop a non-compact analog of
the Verlinde formula, not speaking of the formula (\ref{main}).
Thus, at this stage of the development of the subject considerations
of this paragraph remain mere guesses. However, progress
along these lines may be instrumental in developing a better
technique for integrating over the moduli spaces, and thus,
eventually, for a better understanding of the structure of
string theory.

\bigskip
\noindent{\large \bf Acknowledgements.} L. F. is grateful to
K. Gawedzki for discussions and to the Perimeter Institute for
Theoretical Physics for the support while doing this work. 
K. K. is grateful to R. Roiban for 
discussions and to J. Walcher for reading the manuscript. 
L. F. was supported by CNRS and an ACI-Blanche grant. K. K. was supported by
the NSF grant PHY00-70895. 

\appendix
\section{Some recoupling identities}

The 2-fusion identity:
\be\label{rec-2}
\lower0.4in\hbox{\epsfig{figure=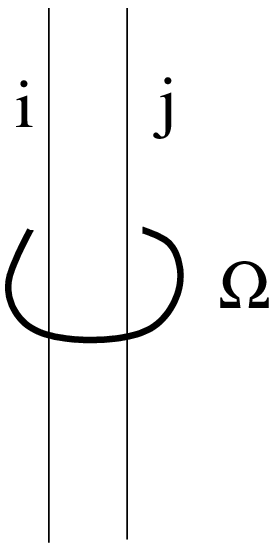,height=0.8in}}=
\delta_{ij} {\eta^{-1}\over {\rm dim}_i}\,\,
\lower0.4in\hbox{\epsfig{figure=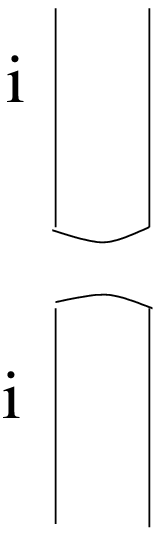,height=0.8in}}
\ee
The 3-fusion identity:
\be\label{rec-3}
\lower0.4in\hbox{\epsfig{figure=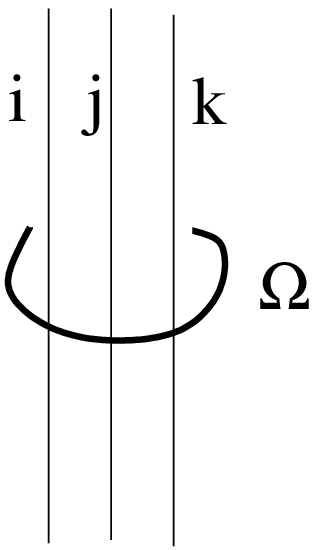,height=0.8in}}=\eta^{-1}\,\,
\lower0.45in\hbox{\epsfig{figure=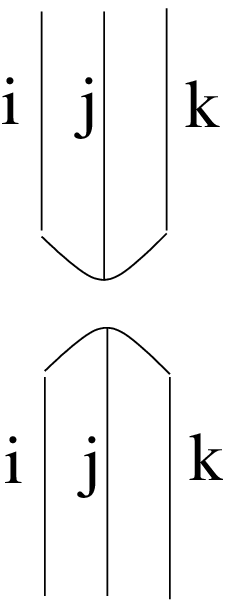,height=0.9in}}
\ee
The 3-vertex is normalized so that:
\be\label{rec-norm}
\lower0.25in\hbox{\epsfig{figure=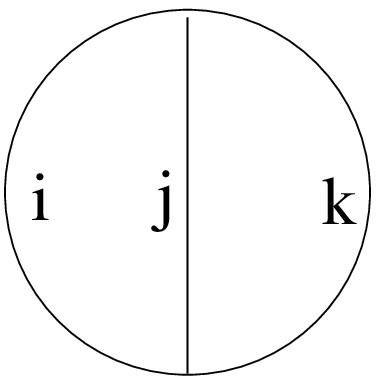,height=0.5in}}=N_{ijk},
\ee
where $N_{ijk}$ is the multiplicity with which the trivial
representation appears in the tensor product of $i, j, k$. For
${\rm SU}(2)$ this is either zero or one.

Another recoupling identity uses the modular S-matrix:
\be\label{rec-s}
\lower0.35in\hbox{\epsfig{figure=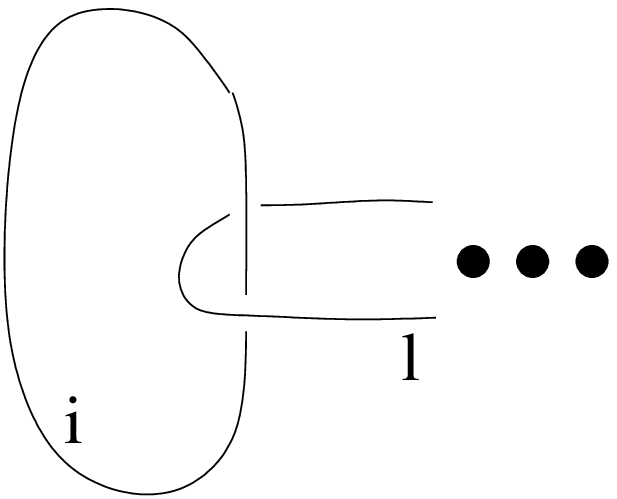,height=0.7in}} =
{S_{il}\over \eta\, {\rm dim}_i {\rm dim}_l}\,\,
\lower0.35in\hbox{\epsfig{figure=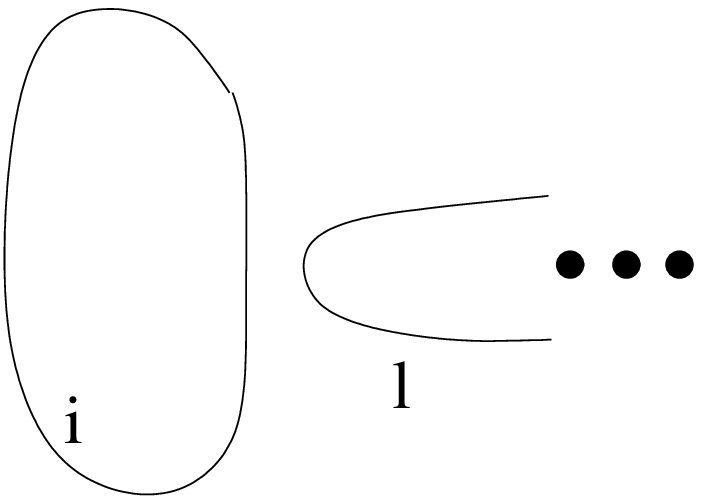,height=0.7in}}
\ee
The dots on the right hand side mean that the open ends can
be connected (in an arbitrary way) to a larger graph.

\section{Proof}

Here we give a proof of the main theorem. 

\bigskip
\noindent{\bf Genus zero case.} We start by working out
the simplest case of the 3-punctured sphere. We choose $\Gamma$ to be
given by a dumbbell. We thus need to compute the following evaluation:
\be\label{fig-1}
\lower0.5in\hbox{\epsfig{figure=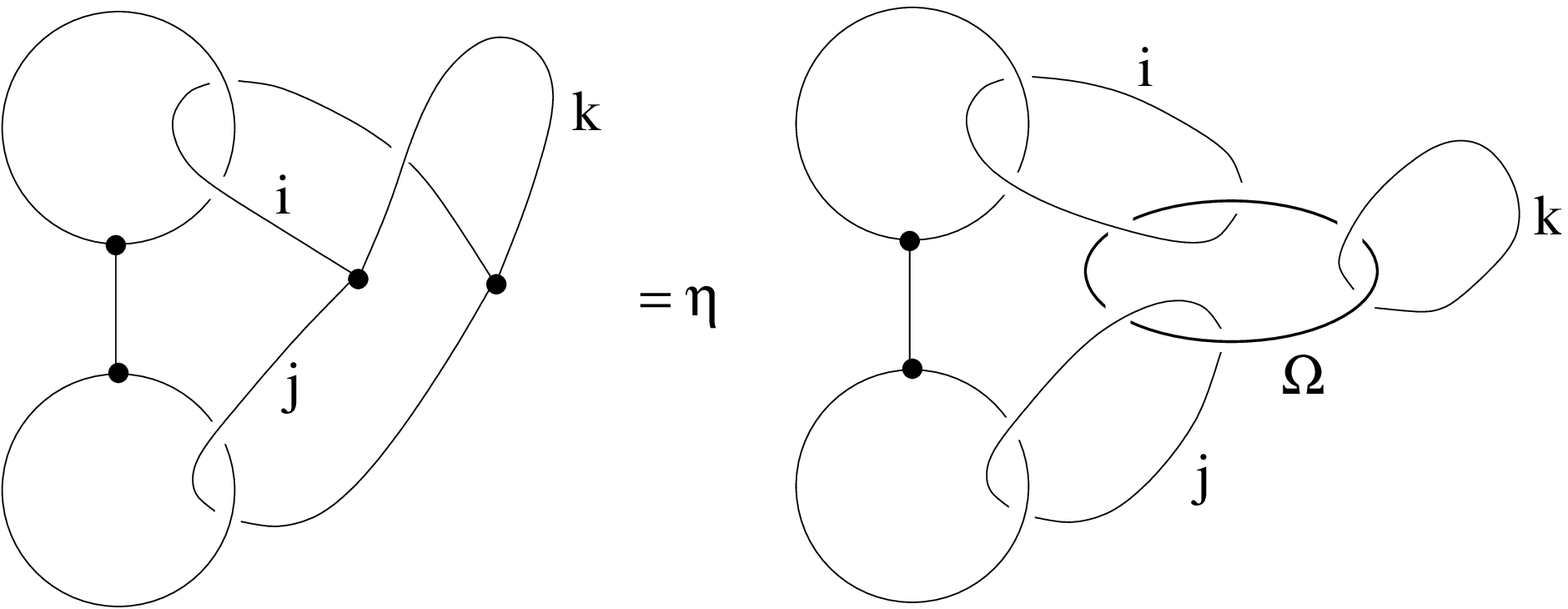,height=1in}}
\ee
Here we have used the observation (\ref{obs}) to replace two
tri-valent vertices of $\Delta\cup\Delta$ by a link with
$\Omega$ inserted. Let us now slide the curve along which $\Omega$
is inserted to go all around the graph $\Gamma$, thus making
one of the curves of the chain-mail $C(\Gamma)$. In the next
step we add two more curves from $C(\Gamma)$ that go around punctures,
and at the same time add two meridian curves with $\Omega$ inserted.
This addition of two pairs of $\Omega$ linked does not change
the evaluation in view of the killing property of $\Omega$.
The steps of sliding the $\Omega$ and adding two new pairs of
curves is shown here:
\be
\lower0.7in\hbox{\epsfig{figure=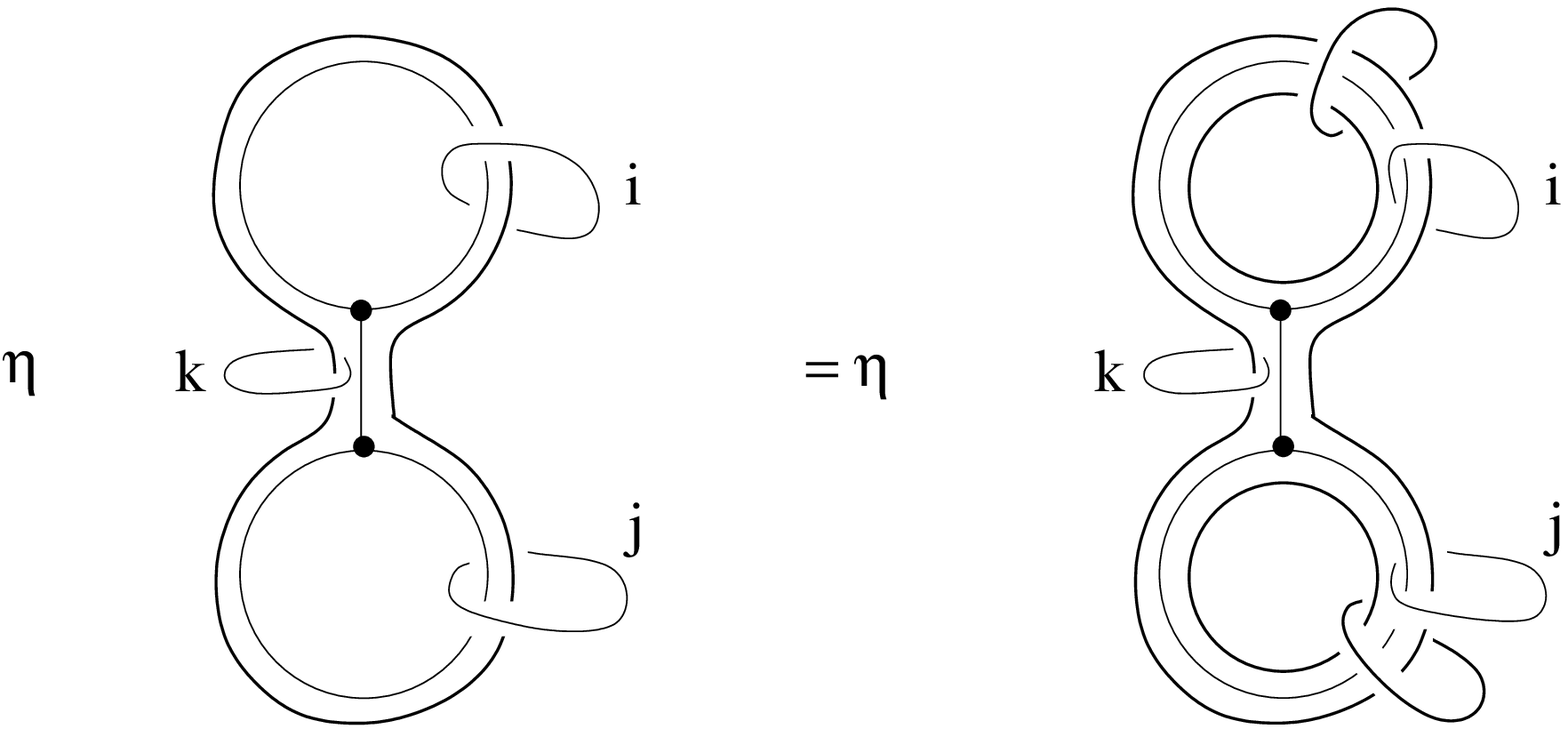,height=1.4in}}
\ee
The last step is to use the sliding property of $\Omega$ to slide
the links labelled $i, j$ inside $\Gamma$:
\be\label{fig-3}
\lower0.7in\hbox{\epsfig{figure=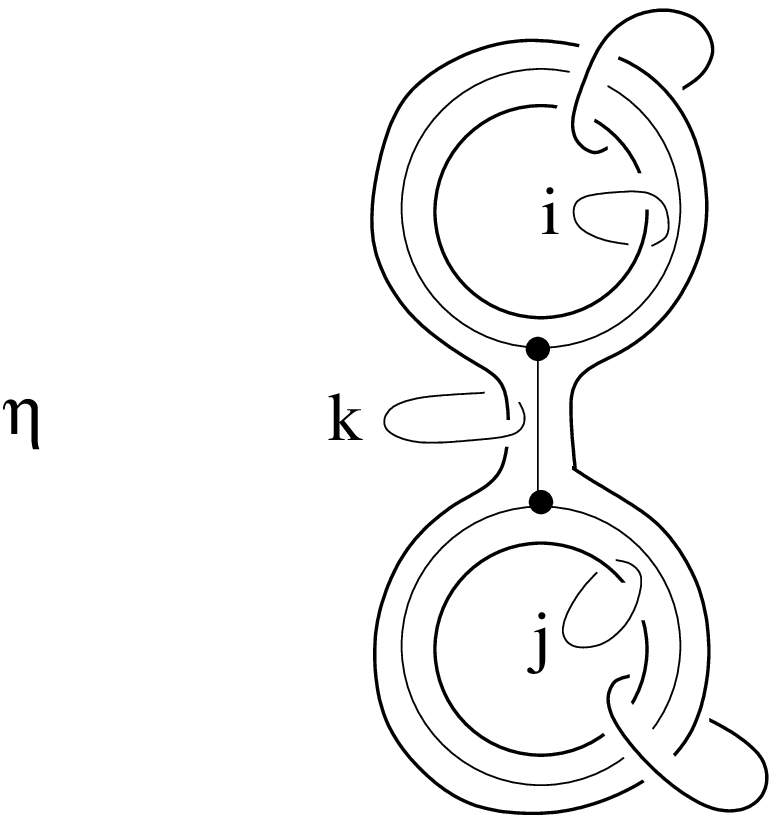,height=1.4in}}
\ee
One can now use the recoupling identity (\ref{rec-s}) to remove
the curves $i, j, k$ at the expense of introducing a factor of
$\eta^{-1} S_{i i'}/{\rm dim}_{i'}$ and similarly for other
loops. Here $i'$ is the representation on the loop from
$C(\Gamma)$ going around the puncture $i$. The element
$\Omega$ on that loop must be expanded (\ref{omega}). The
factor $\eta {\rm dim}_{i'}$ from that expansion is canceling
the factor we got when removing the loop $i$. What is left is
the S-matrix element $S_{i i'}$, with no extra factors.
One can now use the 3-fusion identity (\ref{rec-3}) to
get the formula (\ref{main}). One uses the 3-fusion 2 times,
which produces $\eta^{-2}$. This combines with the factor
of $\eta$ in (\ref{fig-1})-(\ref{fig-3}) to give $\eta^{-1}$,
as prescribed by (\ref{main}) for the case $g=0, n=3$. One can
easily extend this proof to the case $g=0$ arbitrary number of punctures.
To understand the general case, we first find a surgery representation
for $X\times S^1$.

\bigskip
\noindent{\bf Surgery representation for $X\times S^1$.} Let us first
understand the genus one case. A surgery representation for $X_{1,1}\times S^1$
is given by the following link:
\be\label{surgery}
\lower0.35in\hbox{\epsfig{figure=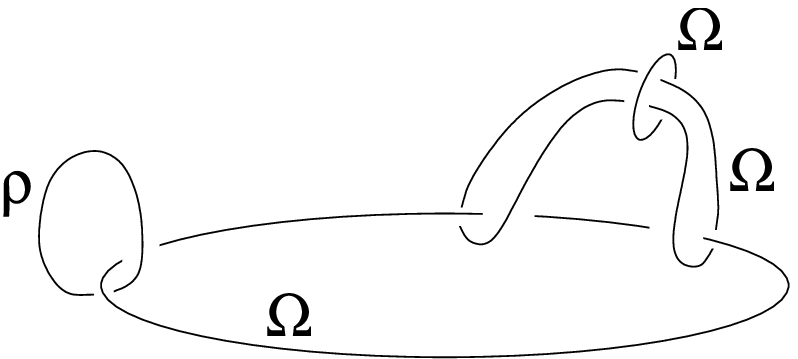,height=0.7in}}
\ee
One must insert the element $\Omega$ into all components, and
evaluate in $S^3$. Representing all the $\Omega$'s as the
sum (\ref{omega}) and using the recoupling identity (\ref{rec-s})
it is easy to show that (\ref{surgery}) gives the correct
expression $\eta I(L) = \sum_{\rho'} S_{\rho \rho'}/S_{0\rho'}$ 
for the dimension. The same surgery representation was noticed in 
\cite{G}.The generalization to higher
genus and to a larger number of punctures is straightforward. It
is given by the following link:
\be\label{surgery-general}
\lower0.6in\hbox{\epsfig{figure=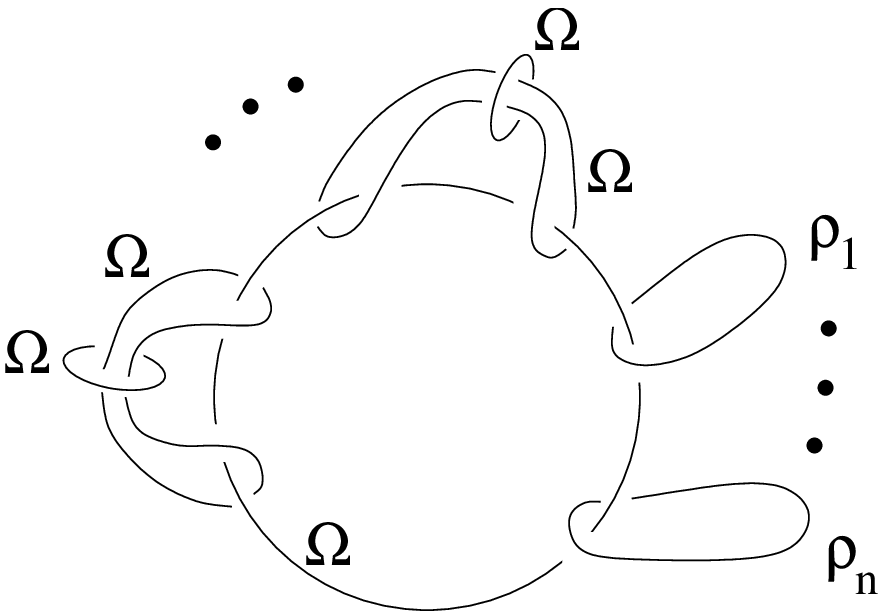,height=1.2in}}
\ee

\bigskip
\noindent{\bf General case.} We will work out only the (1,1) case.
General case is treated similarly. We first note that the formula
(\ref{main}) for (1,1) case can be obtained as the result of the
following evaluation:
\be\label{case-1}
\lower0.5in\hbox{\epsfig{figure=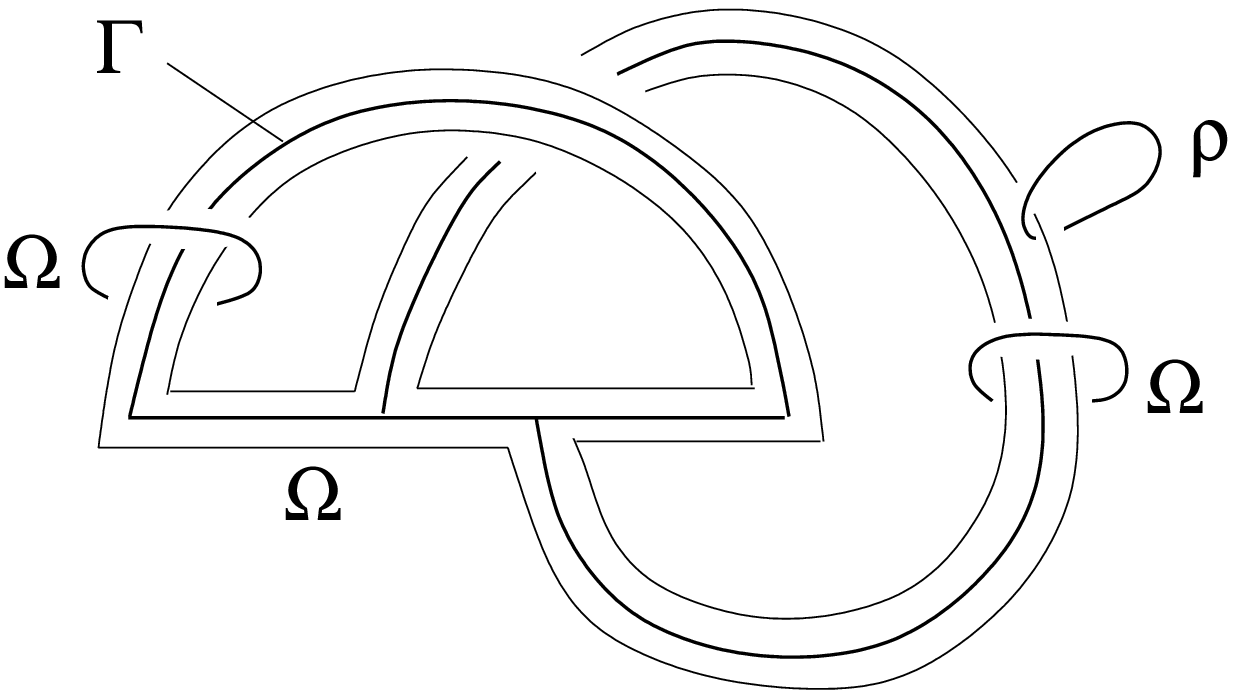,height=1in}}
\ee
This link is to be evaluated in $S^3$ and, as usual, the result multiplied 
by the factor of $\eta$. This gives (\ref{main}) specialized to the
case (1,1). It is now a matter of patience to verify that by the
isotopy moves in $S^3$ the above link can be brought to the form:
\be\label{case-2}
\lower0.5in\hbox{\epsfig{figure=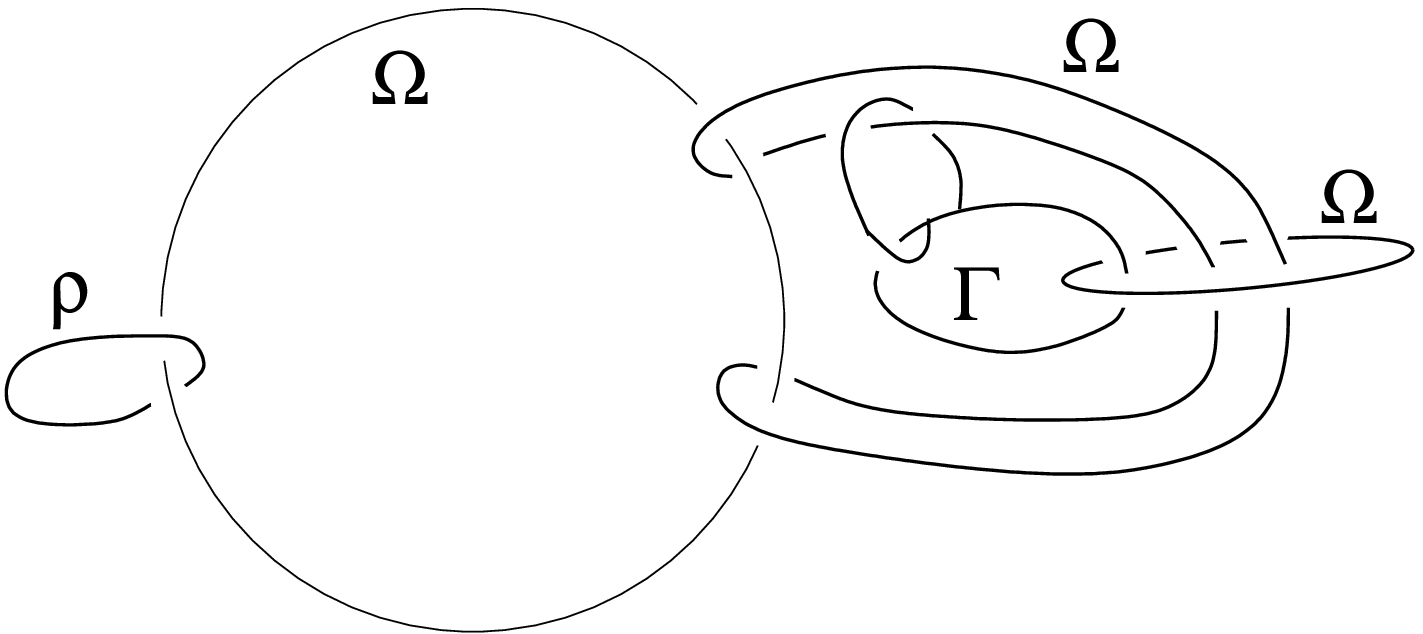,height=1in}}
\ee
This is the correct surgery representation for $X_{1,1}\times S^1$
with the graph $\Gamma$ inside. Thus, (\ref{main}) indeed gives
the evaluation $I(X\times S^1, \Gamma^\psi)$, which, in view
of (\ref{part-spin-net}), proves the theorem.

\newcommand{\hep}[1]{{\tt hep-th/{#1}}}
\newcommand{\gr}[1]{{\tt gr-qc/{#1}}}


\begin{thebibliography}{99}

\bibitem{FK} J.\ Frohlich and C.\ King, Two-dimensional conformal
field theory and three-dimensional topology, {\sl Int.\ J.\ Mod.\ Phys.\ }
{\bf A4} 5321 (1989).

\bibitem{Schweigert} C.\ Schweigert, J.\ Fuchs and J.\ Walcher,
Conformal field theory, boundary conditions and applications to
string theory, \hep{0011109}.

\bibitem{Turaev-Book} V.\ G.\ Turaev, Quantum invariants of
knots and 3-manifolds, de Gruyter Studies in Mathematics, {\bf 18},
New York, 1994.

\bibitem{Verlinde-H} H.\ Verlinde,
Conformal field theory, two-dimensional quantum
gravity and quantization of Teichmuller space, {\sl Nucl.\ Phys.\ }
{\bf B337} 652-680 (1990).

\bibitem{Teich} K.\ Krasnov, On holomorphic factorization in
asymptotically AdS 3d gravity, \hep{0109198}.

\bibitem{TV} V.\ G.\ Turaev and O.\ Yu.\ Viro, State-sum invariants of
3-manifolds and quantum 6j-symbols, {\sl Topology} {\bf 31} 865-902
(1992).

\bibitem{Gawedzki} K.\ Gawedzki, I.\ Todorov, P.\ Tran-Ngoc-Bich,
Canonical quantization of the boundary Wess-Zumino-Witten model, \hep{0101170}.

\bibitem{Schw-1} J.\ Fuchs, I.\ Runkel and C.\ Schweigert,
TFT construction of RCFT correlators I: Partition functions,
\hep{0204148}.

\bibitem{BP} R.\ Benedetti and C.\ Petronio, On Roberts' proof of
the Turaev-Walker theorem, {\sl J.\ Knot Theory and
Ramifications} {\bf 5} 427-439 (1996).

\bibitem{Witten-Jones} E.\ Witten, Quantum field theory and the Jones
polynomial, {\sl Comm.\ Math.\ Phys.\ } {\bf 121} 351 (1989).

\bibitem{Goldman} W.\ Goldman, The symplectic nature of fundamental
groups of surfaces, {\sl Adv.\ Math.\ } {\bf 54} 200-225 (1984).

\bibitem{Witten-Holomorphic} E.\ Witten, On holomorphic factorization of
WZW and coset models, {\sl Comm.\ Math.\ Phys.\ }{\bf 144} 189-212 (1992).

\bibitem{Verlinde-E} E.\ Verlinde, Fusion rules and modular transformations
in two-dimensional conformal field theory, {\sl Nucl.\ Phys.\ }{\bf B 300}
360-376 (1988).

\bibitem{MS} G.\ Moore and N.\ Seiberg, Classical and quantum
conformal field theory, {\sl Comm.\ Math.\ Phys.\ }{\bf 123}
177-254 (1989).

\bibitem{Baez} J.\ Baez, Spin networks in nonperturbative quantum gravity,
{\sl in The Interface of Knots and Physics}, ed. Louis Kauffman, A.M.S.,
Providence, 1996, pp. 167-203.

\bibitem{Penner} R.\ C.\ Penner, The decorated Teichmuller space
of punctured surfaces, {\sl Comm.\ Math.\ Phys.\ }{\bf 113} 299-339 (1987).

\bibitem{Roberts} J.\ Roberts, Skein theory and Turaev-Viro invariants,
{\sl Topology} {\bf Vol. 34} 771-787 (1995).

\bibitem{Turaev} V.\ Turaev, Quantum invariants of links and 3-valent
graphs in 3-manifolds, {\sl Publications Mathematiques d' IHES},
{\bf No. 77} 121-171 (1993).

\bibitem{RT} N.\ Y.\ Reshetikhin and V.\ G.\ Turaev, Invariants
of 3-manifolds via link polynomials and quantum groups, {\sl Invent.\
Math.\ }{\bf 103} 547-597 (1991).

\bibitem{KL} L.\ Kauffman and S.\ Lins, Temperley-Lieb recoupling theory
and invariants of 3-manifolds, Princeton University Press, Princeton, 1994.

\bibitem{Boulatov} D.\ V.\ Boulatov, Q QCD in 2 dimensions and G/G model,
{\sl Mod.\ Phys.\ Lett.\ }{\bf A8} 3491-3496 (1993).

\bibitem{Fock} V.\ V.\ Fock and A.\ A.\ Rosly, Poisson structure on moduli
space of flat connections on Riemann surfaces and $r$-matrix,
{\tt math.QA/9802054}.

\bibitem{Alekseev} A.\ Yu.\ Alekseev, Integrability in the Hamiltonian
Chern-Simons theory, {\sl St.\ Petersburg Math.\ J.\ }{\bf 6} 241-253 (1995),
\hep{9311074}.

\bibitem{AM} A.\ Yu.\ Alekseev and A.\ Z.\ Malkin, Symplectic structure
on the moduli space of flat connections on a Riemann surface,
{\sl Comm.\ Math.\ Phys.\ }{\bf 166} 99-120 (1995), \hep{9312004}.

\bibitem{AS} A.\ Yu.\ Alekseev and V.\ Shomerus, Representation theory
of Chern-Simons observables, {\tt q-alg/9503016}.

\bibitem{AGS-1} A.\ Yu.\ Alekseev, H.\ Grosse and V.\ Shomerus,
Combinatorial quantization of the Hamiltonian Chern-Simons theory I,
{\sl Comm.\ Math.\ Phys.\ }{\bf 172} 317-358 (1995), \hep{9403066}.

\bibitem{AGS-2} A.\ Yu.\ Alekseev, H.\ Grosse and V.\ Shomerus,
Combinatorial quantization of the Hamiltonian Chern-Simons theory II,
{\sl Comm.\ Math.\ Phys.\ }{\bf 174} 561-604 (1995), \hep{9408097}.

\bibitem{BNR} E.\ Buffenoir, K.\ Nori and Ph.\ Roche, Hamiltonian
quantization of Chern-Simons theory with $SL(2,\C)$ gauge group,
\hep{0202121}.

\bibitem{FS} D.\ Friedan and S.\ Shenker, The analytic geometry of
2-dimensional conformal field theory,
{\sl Nucl.\ Phys.\ }{\bf B 281} 509-545 (1987).

\bibitem{PR} G.\ Ponzano and T.\ Regge, Semiclassical limit of Racah
coefficients, in: {\sl Spectroscopic and group theoretic methods in
physics}, ed. F.\ Block et. al., North Holland, Amsterdam, 1968.

\bibitem{MT} S.\ Mizoguchi and T.\ Tada, 3-dimensional Gravity
from the Turaev-Viro Invariant, {\sl Phys.\ Rev.\ Lett.\ } {\bf 68}
1795-1798 (1992).

\bibitem{Fock-1} V. V.\ Fock, Dual Teichmuller spaces, {\tt dg-ga/9702018};

L.\ Chekhov and V. V.\ Fock, Quantum Teichmuller space,
{\sl Theor.\ Math.\ Phys.\ }{\bf 120} 1245-1259 (1999),
also available as {\tt math/9908165}.

\bibitem{Kashaev} R. M.\ Kashaev, Quantization of Teichmuller spaces and the
quantum dilogarithm, available as {\tt q-alg/9705021}.

\bibitem{Teschner} B.\ Ponsot and J.\ Teschner, Liouville bootstrap
via harmonic analysis on a non-compact quantum group, \hep{9911110}.

\bibitem{G} E.\ Guadagnini and L.\ Pilo, Three-dimensional topology
and Verlinde formula, {\sl Nucl.\ Phys.\ }{\bf B433} 597-624 (1995).

\end{thebibliography}
\end{document}